\documentclass[12pt]{iopart}

\expandafter\let\csname equation*\endcsname\relax
  \expandafter\let\csname endequation*\endcsname\relax
  \usepackage{amsmath}

\usepackage{amssymb}
\usepackage[usenames,dvipsnames]{color} 
\usepackage{epsfig}
\usepackage{dsfont}
\usepackage{float}
\usepackage{hyperref}
\usepackage{multirow}
\usepackage{hhline}

\hypersetup{
    colorlinks=true,
	linkcolor=Blue,
	citecolor=Blue,
	urlcolor=Black
}

\newcommand\smallO[1]{
  \mathchoice
    {{\scriptstyle\mathcal{O}}}
    {{\scriptstyle\mathcal{O}}}
    {{\scriptscriptstyle\mathcal{O}}}
    {\scalebox{.7}{$\scriptscriptstyle\mathcal{O}$}}
  {\left(#1\right)}}

\newcommand{\req}[1]{Eq.~\ref{#1}}

\DeclareMathOperator{\arccot}{arccot}
\newcommand{\sgn}{\text{sgn}}

\usepackage{upquote,fancyvrb,subfig,microtype,sistyle,refstyle,wasysym}
\usepackage[process=auto,crop=pdfcrop]{pstool}

\begin{document}


\title{Algebraic synthesis of time-optimal unitaries \\ in $SU(2)$ with alternating controls}

\author{Clarice D.~Aiello$^{1,2,3}$, Michele Allegra$^{2,4,5,\star}$, B\"orge Hemmerling$^{3,6,\star}$, Xiaoting Wang$^{2}$ and Paola Cappellaro$^{2,3,7}$}

\ead{clarice@mit.edu}

\address{$^1$ Department of Electrical Engineering and Computer Science, \\ Massachusetts Institute of Technology, Cambridge, Massachusetts 02139, USA}
\address{$^2$ Research Laboratory of Electronics, \\ Massachusetts Institute of Technology, Cambridge, Massachusetts 02139, USA}
\address{$^3$ Center for Ultracold Atoms, \\ Massachusetts Institute of Technology, Cambridge, Massachusetts 02139, USA, \\ and Harvard University, Cambridge, Massachusetts 02138, USA} 
\address{$^4$ Dipartimento di Fisica, Universit\`a di Torino \& INFN, \\ Sezione di Torino,
I-10125, Torino, Italy}
\address{$^5$ Institute for Scientific Interchange Foundation, I-10126 Torino, Italy} 
\address{$^6$ Department of Physics, Harvard University, Cambridge, Massachusetts 02138, USA}
\address{$^7$ Department of Nuclear Science and Engineering, \\ Massachusetts Institute of Technology, Cambridge, Massachusetts 02139, USA}
\address{$^\star$ These authors contributed equally to this work.}

\begin{abstract}
We present an algebraic framework to study the time-optimal synthesis of arbitrary unitaries in $SU(2)$, when the control set is restricted to 
 rotations around two non-parallel axes in the Bloch sphere. Our method bypasses commonly used control-theoretical techniques, and easily imposes necessary conditions on time-optimal sequences. In a straightforward fashion, we prove that time-optimal sequences are solely parametrized by three rotation angles and derive general bounds on those angles as a function of the relative rotation speed of each control and the angle between the axes. 
 Results are substantially different whether both clockwise and counterclockwise rotations about the given axes are allowed, or only clockwise rotations. In the first case, we prove that any finite time-optimal sequence is composed at most of five control concatenations, while for the more restrictive case, we present scaling laws on the maximum length of any finite time-optimal sequence. 
The bounds we find for both cases are stricter than previously published ones and severely constrain the structure of time-optimal sequences, allowing for an efficient numerical search of the time-optimal solution. 
Our results can be used to find the time-optimal evolution of qubit systems under the action of the considered control set, and thus potentially increase the number of realizable unitaries before decoherence. 
\end{abstract}

\maketitle
\tableofcontents


\section{Introduction}

The power of many quantum-enabled technologies, especially quantum computing, critically depends on the possibility of implementing an algorithm before the quantum system has decohered. Given constraints in the control fields, it is hence desirable to implement unitaries (or gates) in the 
shortest possible time.  While time-optimal control has been often studied in the state-to-state
framework, unitary gate generation, or synthesis, is of even greater relevance in that it can be incorporated into control protocols regardless of the initial state of the evolving quantum system. 

Time-optimal unitary synthesis in $SU(2)$ has been studied in the context of a continuous control set composed of rotations around any axis in a plane in the Bloch sphere~\cite{Khaneja01,Boozer12,Garon13}. 

In this work, we address the challenge of synthesizing any $SU(2)$ unitary $U_{\scriptsize{\textrm{goal}}}$ in a time-optimal way using a control set only composed of alternating rotations $\mathsf{X},\mathsf{V}$ around two non-parallel axes in the Bloch sphere; that is, we study the unitary synthesis via the concatenation
\begin{equation}
U_{\scriptsize{\textrm{goal}}} = \mathsf{X}(t_n) \cdot \ ... \ \cdot \mathsf{V}(t_2) \cdot \mathsf{X}(t_1) \cdot \mathds{1} \ .
\end{equation}
 
The discrete control set of interest here has mostly been studied in a state-to-state transfer framework for bounded controls that can vary in magnitude~\cite{Boscain05, Boscain06, Boscain12}, where it emerges as the time-optimal solution. 
It is experimentally relevant in quantum systems for which amplitude and phase modulation of the control fields is relatively difficult; and in systems with restricted control degrees of freedom. An example of the latter is a nuclear $^{13}$C spin hyperfine-coupled to the electronic spin of nitrogen-vacancy (NV) centers in diamond~\cite{Jelezko05, Childress06}. Due to the anisotropy in the hyperfine coupling, the electronic spin can be regarded as an actuator~\cite{Hodges08}; its switching between spin states steers the nuclear spin evolution, thus providing an alternative to the slow and noisy radio-frequency addressing of the $^{13}$C. Moreover, the same control set is also relevant for robotics and satellite motion in $SO(3)$~\cite{Boscain05,Trelat12}, due to the two-to-one homomorphic mapping of $SU(2)$ onto $SO(3)$.
 
The standard approach to time-optimal control usually involves general, but rather abstract optimization protocols, such as the
Pontryagin maximum principle~\cite{Pontryagin61}, or variational~\cite{Carlini06} and geometric control methods~\cite{Khaneja01, Boscain06, Boscain12}, all of which are hard to use in practice to find solutions for specific cases. 
In the case of dynamics generated by a smoothly-varying Hamiltonian, a combination of optimization and geometric techniques lead to a simple characterization of the time-optimal solutions~\cite{Boozer12,Garon13}. 
In the case of alternating controls, though, such methods either fail because of non-smooth changes in the Hamiltonian, or become convoluted in all but some specific cases, thereby losing in generality. Besides, numerical methods to find the time-optimal solution in this case usually rely on the integration of rather involved systems of differential equations. Driven by experimental needs, we take a different approach, and use only algebraic methods first developed in~\cite{Billig13} that turn out to be more powerful than more refined mathematical techniques, at least for the problem at hand.
We obtain fully general results for the structure of time-optimal sequences in $SU(2)$, which can then be exploited to boost the efficiency of a numerical search. 

This paper is organized as follows. After clarifying both the precise problem we tackle in this work and the related notation in Section~\ref{subs:statement}, we proceed by deriving our main results in Section~\ref{sec:results}. These consist in the necessary characteristics of time-optimal concatenations of control elements generating any $SU(2)$ unitary, and impose bounds on: the maximum number of independent parameters, namely three rotation angles; their values; and the maximal concatenation length. A full summary of our results is presented in the three Tables of Subsection~\ref{subs:summary}, which can be used as a reference,  independently of the preceding mathematical derivation of results. Finally, in Section~\ref{sec:apps}, we discuss experimental settings for which 
the driving of qubits according to time-optimal controls which are numerically found using our method might prove beneficial. A conclusion follows in Section~\ref{sec:conc}.
\section{Statement of problem and notation}
\label{subs:statement}

\begin{figure}[t]
\centering
\includegraphics[width=0.25\columnwidth]{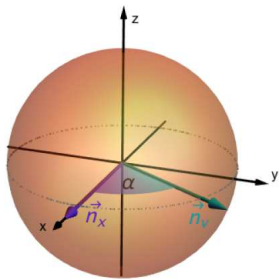}
\caption{We investigate the generation of any $SU(2)$ element by solely allowing rotations around two non-parallel axis in the Bloch sphere, namely $\vec{n}_x$ and $\vec{n}_v$, which are separated by an angle $\alpha$.}
\label{fig:0}
\end{figure}

We investigate the time-optimal synthesis of $SU(2)$ elements up to a global phase, using an alternating control set denoted by $\{ \mathsf{X}(t_x) \equiv e^{-i\frac{t_x}{2}\sigma_x}, \mathsf{V}(t_v) \equiv e^{-i\frac{t_v}{2}\sigma_v} \}$. Here, $\sigma_v = \cos(\alpha)\sigma_x + \sin(\alpha)\sigma_y$, with $\alpha \in \ ]0,\pi[$ and $\sigma_{x,y}$ the Pauli matrices; $\alpha$ is usually fixed by experimental constraints. The controls represent rotations of angle $t_{x,v}$ around two axes in the Bloch sphere parametrized by $\vec{n}_x = (1,0,0)$ and $\vec{n}_v = (\cos(\alpha),\sin(\alpha),0)$, and separated by an angle $\alpha$. This situation is depicted in Figure~\ref{fig:0}. For $\alpha = \pi/2$, the controls are orthogonal and $\mathsf{V}(\cdot) = \mathsf{Y}(\cdot)$, with $\mathsf{Y}(t_y) \equiv e^{-i\frac{t_y}{2}\sigma_y}$. 

This restricted control set confers complete controllability in $SU(2)$ up to a global phase, $\forall \ \alpha \neq  0,\pi$~\cite{Jurdjevic72}; moreover, any element of $SU(2)$ can be generated by the control set (albeit in a non-time-optimal way) in at most $(\lfloor\frac{\pi}{\alpha}\rfloor + 2)$ concatenations~\cite{Lowenthal71}, where $\lfloor \ \rfloor$ indicates the integer part. 

Experimental constraints determine whether rotations can be realized only in the clockwise direction or in both clockwise and counterclockwise directions. In the first case we have  $t_{x,v} \in \ ]0, 2\pi[ $, with two accessible Hamiltonians, $\{\sigma_x, \sigma_v\}$; in the second case we can either consider four Hamiltonians, $\{\pm\sigma_x,\pm\sigma_v\}$ with $t_{x,v} \in\ ]0,\pi]$, or, equivalently, two Hamiltonians with $t_{x,v} \in \ ]\!-\pi, \pi]$, as we will do in the following. Our analysis is subdivided accordingly,  in cases noted $t > 0$ and $t \lessgtr 0$. 

Additionally, in physical realizations, it is often the case that rotations around distinct axes  have different evolution speeds. To account for that, we introduce a dimensionless parameter $\kappa \in [0, 1]$ and assume, without loss of generality, that a rotation $\mathsf{V}(t_v)$ is effectively synthesized in a (shorter or equal) time $\kappa |t_v|$.  

We call `$n$-sequence' the synthesis of a unitary $U_{\scriptsize{\textrm{goal}}}$ using $n$ alternating controls. An $n$-sequence is time-optimal if it has minimum time cost among sequences of all lengths generating $U_{\scriptsize{\textrm{goal}}}$. A time-optimal sequence can be of finite length or infinite. It is immediate that any subsequence of a time-optimal sequence must be time-optimal itself. In the text, we denote such subsequences   $U^\star$, as in $U_{\scriptsize{\textrm{goal}}} = ...\ \cdot \ U^\star \ \cdot \ ... \cdot  \ \mathds{1}$.

In what follows, we present necessary conditions that time-optimal sequences generating any $U_{\scriptsize{\textrm{goal}}} \in SU(2)$ must obey.


\section{Results}
\label{sec:results}

\subsection{Relationship between internal rotation angles}
\label{subs:relationship}
The problem of finding a time-optimal sequence seems at first intractable since it requires optimizing over a large -- possibly infinite -- number of parameters. Here we show instead that three angles are sufficient to parametrize time-optimal sequences of any length (both finite and infinite). 
Our proof generalizes and strengthens previous results~\cite{Billig13} that were restricted to the case of clockwise rotations, and that were derived through a limited critical-point analysis involving only the first derivative, but not higher derivatives; the latter, as we show next, allow for a much more thorough characterization of critical points. 

The starting point of the analysis is a perturbative approach that fixes the relationship between rotation angles in any time-optimal sequence of length $n \geq 4$~\cite{Billig13} (sequences with $n\leq3$ are trivially parametrized by at most three angles). Assuming the 4-subsequence
\begin{equation}
U^\star = \mathsf{X}(t_f) \cdot \mathsf{V}(t_v) \cdot \mathsf{X}(t_x) \cdot \mathsf{V}(t_i) 
\label{eq2}
\end{equation}
is time-optimal, the total time needed to synthesize $U^\star$, $\mathcal{T} \equiv |t_f| + \kappa|t_v| + |t_x| + \kappa|t_i|$, is at a global minimum. 

Let all times in Eq.~\ref{eq2} depend on a parameter $\delta$, so that $t = t(\delta)$. We examine an infinitesimal perturbation of the sequence
\begin{equation}
\mathsf{X}(t_f(\delta)) \cdot \mathsf{V}(t_v(\delta)) \cdot \mathsf{X}(t_x(\delta)) \cdot \mathsf{V}(t_i(\delta))=U^\star + dU^\star + \smallO{\delta^2}
\end{equation} 
that keeps the unitary unchanged to first order, $dU^\star = 0$. 
By expanding the unitaries to first order in $\delta$ around zero,
\begin{equation}
\mathsf{X}(t(\delta)) \approx \mathsf{X}(t(0)) \cdot \left(\mathds{1} - i \ \frac{\sigma_x}{2}\cdot \delta \left. \frac{dt}{d\delta}\right|_{\delta = 0}\right) \equiv \mathsf{X}(t(0)) \cdot \left(\mathds{1} - i \ \frac{\sigma_x}{2}\cdot \epsilon \right) \ ,
\end{equation}
where we have defined $\delta \left. \frac{dt}{d\delta}\right|_{\delta = 0} \equiv \epsilon$;
and by using relationships such as 
\begin{align}
\sigma_x \cdot \mathsf{V}(t_v) &= \mathsf{V}(t_v)\cdot \mathsf{V}(-t_v)\cdot \sigma_x \cdot \mathsf{V}(t_v) \nonumber \\
&= \mathsf{V}(t_v)\cdot \left( \cos^2\left(\frac{t_v}{2}\right) \sigma_x  + \sin^2\left(\frac{t_v}{2}\right) \sigma_v\sigma_x\sigma_v - i\sin\left(\frac{t_v}{2}\right)\cos\left(\frac{t_v}{2}\right)[\sigma_x, \sigma_v]    \right)\nonumber \\
 &\equiv \mathsf{V}(t_v)\cdot \eta \ ,
\end{align}
and similarly
\begin{equation}
\mathsf{X}(t_x) \cdot \sigma_v  \equiv \eta' \cdot \mathsf{X}(t_x) \ ,
\end{equation}
with
\begin{equation}
\eta' \equiv \left( \cos^2\left(\frac{t_x}{2}\right) \sigma_v  + \sin^2\left(\frac{t_x}{2}\right) \sigma_x\sigma_v\sigma_x - i\sin\left(\frac{t_x}{2}\right)\cos\left(\frac{t_x}{2}\right)[\sigma_v, \sigma_x]    \right) \ ,
\end{equation}
we find that
\begin{equation}
U^\star + dU^\star = \mathsf{X}(t_f)\cdot \mathsf{V}(t_v)\cdot\left( \mathds{1}- i\frac{\epsilon_f}{2}\eta \right)\cdot \left( \mathds{1} - i\frac{\epsilon_v}{2}\sigma_v \right)\cdot\left( \mathds{1} - i\frac{\epsilon_x}{2}\sigma_x \right)\cdot \left( \mathds{1}- i\frac{\epsilon_i}{2}\eta' \right) \cdot\mathsf{X}(t_x)\cdot \mathsf{V}(t_i) \ .
\end{equation}
Imposing $dU^\star = 0$ gives
\begin{equation}
\label{eqsy}
\left( \mathds{1}- i\frac{\epsilon_f}{2}\eta \right ) \cdot \left( \mathds{1} - i\frac{\epsilon_v}{2}\sigma_v \right)\cdot \left( \mathds{1} - i\frac{\epsilon_x}{2}\sigma_x \right)\cdot \left( \mathds{1}- i\frac{\epsilon_i}{2}\eta' \right ) = \mathds{1} \ .
\end{equation}
To first order in $\epsilon_{f,v,x,i}$,~\req{eqsy} yields three independent constraint equations, linear in $\epsilon_{f,v,x,i}$.  
In addition, by assumption of time-optimality, the first derivative of the total time must obey, for $\delta \neq 0$,
 \begin{align}
\delta \frac{d\mathcal{T}}{d\delta} &= \sgn(t_f) \cdot \delta \left. \frac{dt_f}{d\delta}\right|_{\delta = 0} + \kappa \ \sgn(t_v) \cdot \delta \left. \frac{dt_v}{d\delta}\right|_{\delta = 0}  \nonumber \\ \ & + \sgn(t_x)\cdot \delta \left. \frac{dt_x}{d\delta}\right|_{\delta = 0} + \kappa \ \sgn(t_i)\cdot \delta \left. \frac{dt_i}{d\delta}\right|_{\delta = 0} \nonumber \\ 
&= \sgn(t_f) \ \epsilon_f + \kappa \ \sgn(t_v) \ \epsilon_v + \sgn(t_x)\ \epsilon_x + \kappa \ \sgn(t_i)\ \epsilon_i  \nonumber  \\ 
&= 0 \ .
\end{align} 


Solutions of the above four equations give $t_{v}$ as a function of $t_{x}$ (or vice-versa) and are obtained upon imposing the non-triviality condition given by 
\vspace{0.1cm}
\begin{equation}
\label{eq:cst_eqs1}
\left| \begin{array}{cccc}
\cos(t_v) & 0 & 1 & 2\cos(\alpha)\sin^2\left(\frac{t_x}{2}\right)\\
2\cos(\alpha)\sin^2\left(\frac{t_v}{2}\right) & 1 & 0 & \cos(t_x)\\
\sin(t_v) & 0 & 0 & \sin(t_x)\\
\sgn(t_f) & \kappa \ \sgn(t_v) & \sgn(t_x)  & \kappa \ \sgn(t_i)
\end{array} \right| = 0 \ .
\end{equation}
Analogous calculations provide similar constraint equations for a 4-subsequence of type $U^\star =  \mathsf{V}(t_f) \cdot \mathsf{X}(t_x) \cdot \mathsf{V}(t_v) \cdot \mathsf{X}(t_i)$. 
General solutions to~\req{eq:cst_eqs1} are found by considering the eight relative sign combinations for $\{t_i, t_x, t_v, t_f\}$; they fix $t_v$ as a function of $t_x$ for both finite $n \geq 4$ and infinitely long time-optimal sequences. These solutions are presented in Table~\ref{tab:sol_cst_eqs}. In this table, the sign vector entry corresponds to the signs of $\{t_i, t_x, t_v, t_f\}$. 

Importantly, in true minima, the second derivative of the total time function must obey $\frac{d^2\mathcal{T}}{d\delta^2} > 0$, independently of the perturbation parameter $\delta$. Note that, in~\cite{Billig13}, only the criticality condition $\frac{d\mathcal{T}}{d\delta} = 0$ is considered.
To discriminate the true minima, we perform a calculation similar to the preceding one, but expanding to second order in $\delta$ around $\delta = 0$, thereby obtaining:
\begin{align}
\delta^2\frac{d^2\mathcal{T}}{d\delta^2} &= \sgn(t_f) \cdot \delta^2 \left. \frac{d^2t_f}{d\delta^2}\right|_{\delta = 0} + \kappa \ \sgn(t_v) \cdot \delta^2 \left. \frac{d^2t_v}{d\delta^2}\right|_{\delta = 0}  \nonumber \\ \ & + \sgn(t_x)\cdot \delta^2 \left. \frac{d^2t_x}{d\delta^2}\right|_{\delta = 0} + \kappa \ \sgn(t_i)\cdot \delta^2 \left. \frac{d^2t_i}{d\delta^2}\right|_{\delta = 0}\ ; \\
\mathsf{X}(t(\delta)) &\approx \mathsf{X}(t(0)) \cdot \left[\mathds{1} - i \ \frac{\sigma_x}{2}\cdot \left(\delta \left. \frac{dt}{d\delta}\right|_{\delta = 0} + \delta^2 \left. \frac{d^2t}{d\delta^2}\right|_{\delta = 0}\right) - \left( \frac{\delta}{2} \left. \frac{dt}{d\delta}\right|_{\delta = 0}\right)^2\right] \ .
\end{align}


All eight cases in Table~\ref{tab:sol_cst_eqs} obey $\frac{d\mathcal{T}}{d\delta} = 0$, but only some of them have regions in $\{\alpha, \kappa, t_x\}$ space with $\frac{d^2\mathcal{T}}{d\delta^2} > 0$. We thus established that if  the experimentally given parameters $\alpha$, $\kappa$ are such that $\frac{d^2\mathcal{T}}{d\delta^2} \ngtr 0$ for all cases (a) through (h), then the time-optimal sequence generating any $U_{\scriptsize{\textrm{goal}}}$ must be $n \leq 3$ long. Note that, if we are restricted to positive times, $\frac{d^2\mathcal{T}}{d\delta^2} \ngtr 0$ is sufficient to ensure that $n \leq 3$.


\begin{table}[hbt]
\centering
\begin{tabular}{|c|c|l|c|c|}
\hline
case&sign&\hspace{1.8cm} relationship $t_v(t_x)$&$\exists \ \delta^2\frac{d^2\mathcal{T}}{d\delta^2} > 0$ & length \\
\hline \hline
(a)&$\{+,+,+,+\}$&$\tan\left(t_v/2\right)  = \tan\left(t_x/2\right) \frac{\kappa - \cos(\alpha)}{1 - \kappa \cos(\alpha)}$ &  yes & $n$,\ $\infty$\\
\hline
(b)&$\{+,+,+,-\}$ &$\tan\left(t_v/2\right) = - \frac{\kappa + \cos(\alpha) + \cos(t_x)(\kappa-\cos(\alpha))}{(1-\kappa \cos(\alpha))\sin(t_x)}$ & yes&$\varnothing$\\
\hline
(c)&$\{+,+,-,+\}$ &$\tan\left(t_x/2\right) =  \frac{1-\kappa\cos(\alpha) + \cos(t_v)(1+\kappa\cos(\alpha))}{(\kappa + \cos(\alpha))\sin(t_v)}$ & yes&$\varnothing$\\
\hline
(d$_1$)&$\{+,+,-,-\}$&$\tan(t_x/2) = - {\kappa}\tan\left(t_v/2\right)$& yes  & $n$\\
\hline
(d$_2$)&$\{+,+,-,-\}$&$ \tan(t_v/2) =\cot(t_x/2)\sec(\alpha) $ & no & $\varnothing$\\
\hline
(e)&$\{+,-,+,+\}$  &$\tan\left(t_v/2\right) = \frac{\kappa - \cos(\alpha) + \cos(t_x)(\kappa+\cos(\alpha))}{(1+\kappa \cos(\alpha))\sin(t_x)}$ & yes &$\varnothing$\\
\hline
(f)&$\{+,-,+,-\}$ &$\tan\left(t_v/2\right)  = -\tan\left(t_x/2\right)  \frac{\kappa + \cos(\alpha)}{1 + \kappa \cos(\alpha)}$ & yes & $\infty$ \\
\hline
(g$_1$)&$\{+,-,-,+\}$& $\tan(t_x/2) =  {\kappa}\tan(t_v/2)$& yes & $n$\\
\hline
(g$_2$)&$\{+,-,-,+\}$&$ \tan(t_v/2) =\cot(t_x/2)\sec(\alpha)$ & no &$\varnothing$\\
\hline
(h)&$\{+,-,-,-\}$ &$\tan\left(t_x/2\right) = -  \frac{1+\kappa\cos(\alpha) + \cos(t_v)(1-\kappa\cos(\alpha))}{(\kappa- \cos(\alpha))\sin(t_v)}$ & yes &$\varnothing$\\
\hline
\end{tabular}
\caption{\label{tab:sol_cst_eqs} Relationship between internal rotation angles in a time-optimal $n$-sequence, $n \geq 4$, and length of candidate time-optimal sequences. Note that case (a) can yield a finite time-optimal sequence only in the $t > 0$ case. Here, the symbol $\varnothing$ indicates sequences that cannot be optimal on the basis of the analysis of Subsection~\ref{subs:bounds}.}
\end{table}

If a time-optimal sequence has length $n > 4$, any 4-subsequence
must be time-optimal. Therefore, all pairs of `internal' rotation angles $\{t_{i}, t_{i+1}\}$, 
with $1 < i < n-1$, must obey the prescribed relations in Table~\ref{tab:sol_cst_eqs}. From this, one immediately
infers that all internal rotation angles $t_{i}$ with $1 < i < n$ are fixed
by a single internal time $t_{x}$.  

In conclusion, time-optimal sequences which are $n \geq 4$ long have only three independent parameters, namely the initial and final rotation angles $t_i, t_f$, and the internal angle parameter $t_x$.  

This simple, yet non-trivial result is the essential keystone
that allows for a numerical search of time-optimal solutions; if this were not the case, in the presence
of a growing number of parameters, any numerical search would soon
become impractical. The numerical analysis will be further simplified
by the results of Subsections~\ref{subs:bounds} and~\ref{outer}, which give additional constraints
of the possible values of the three parameters.

We now summarize the relationship $t_v(t_x)$ by case.  

\vspace{11pt}
\footnotesize $\blacksquare$\normalsize\ \ \underline{\textbf{Case $t > 0$.}} The relationship between internal times for sign combination (a) in Table~\ref{tab:sol_cst_eqs} is~\cite{Billig13}
\begin{equation}
\label{eq:tv_pos}
\tan\left(\frac{t_v}{2}\right) = \tan\left(\frac{t_x}{2}\right) \cdot \frac{\kappa - \cos(\alpha)}{1 - \kappa \cos(\alpha)} \equiv \tan\left(\frac{t_x}{2}\right) \cdot K_1 \ .
\end{equation}
Note that $ \kappa > \cos(\alpha)  \Leftrightarrow K_1  > 0 $ (conversely, $\kappa < \cos(\alpha) \Leftrightarrow K_1 < 0 $). This naturally subdivides case $t > 0$ in two subcases with different structures of time-optimal sequences. 

For $\kappa < \cos(\alpha)$, imposing $\frac{d^2\mathcal{T}}{d\delta^2}$ constrains $t_x < \pi$. 

For $\kappa > \cos(\alpha)$, $\frac{d^2\mathcal{T}}{d\delta^2} > 0$ only holds for $\pi<t_x < \frac{5\pi}{3}$ and $\alpha < \frac{2\pi}{3}$; from this we easily conclude that, if $\alpha > \frac{2\pi}{3}$, finite time-optimal sequences are at most $n = 3$ long. In the limiting case $\kappa = 1$, $t_v = t_x$, with the constraint $t_x > \pi$ imposed by the second derivative condition. 


The relationship of~\req{eq:tv_pos} must be valid for any time-optimal sequence of length $n\geq4$, including for an infinite concatenation of control elements that realizes a given $U_{\scriptsize{\textrm{goal}}}$ in (finite) optimal time. Thus, in this limit, necessarily $t_x \rightarrow 0$, and the relationship between $t_x$ and $t_v$ is obtained by noting that
\begin{equation}
\lim_{t_x \to 0} t_v =(t_x \cdot K_1 + \smallO{t_x})\ \textrm{mod} \ 2\pi \ .
\end{equation}
If $K_1 < 0$, $t_v \rightarrow 2\pi - t_x \cdot |K_1| \approx 2\pi$; an infinite concatenation of control elements in this case would take an infinite time cost. Hence, if $\kappa < \cos(\alpha)$, time-optimal sequences must be finite~\cite{Billig13}. 

Infinite-length time optimal sequences might thus exist only for $\kappa > \cos(\alpha)$. We define a rotation $\mathsf{Q}$ that effectively represents an infinite concatenation of control elements:
\begin{equation}
\mathsf{Q}(t_Q) \equiv e^{-i\frac{t_Q}{2} (\sigma_x + K_1 \sigma_v)} = \lim_{k \to \infty} \left[\mathsf{X}\left(\frac{t_x}{k}\right)\cdot \mathsf{V}\left(\frac{t_v}{k}\right)\right]^k  = \lim_{k \to \infty} \left[\mathsf{V}\left(\frac{t_v}{k}\right)\cdot \mathsf{X}\left(\frac{t_x}{k}\right)\right]^k\ .
\end{equation}
The normalized axis of the $\mathsf{Q}$ rotation, $\vec{n}_q$, exactly bisects $\alpha$ for $\kappa = 1$; as $\kappa$ decreases towards its lower limit $\cos(\alpha)$, the axis approaches $\vec{n}_x$: $(\vec{n}_v \cdot \vec{n}_q) = \kappa \ (\vec{n}_x \cdot \vec{n}_q)$. The axis normalization is given by
\begin{equation}
N_q \equiv \frac{\sqrt{1+\kappa^2-2\kappa\cos(\alpha)}\sin(\alpha)}{1-\kappa\cos(\alpha)} \ .
\end{equation}
The implementation time cost associated with such a $\mathsf{Q}(t_Q)$ rotation is $t_Q \left(1 + \kappa \cdot K_1 \right)$, where $t_Q \in \ \left]0, \frac{2\pi}{N_q} \right[ \ $ is in principle unbounded. For simplicity, we define the renormalized time $t_q \equiv t_Q \cdot N_q$, which is bounded as $t_q \in \ ]0, 2\pi[ \ $. 

\vspace{11pt}
\footnotesize $\blacksquare$\normalsize\ \ \underline{\textbf{Case $t \gtrless 0$.}} All eight relative sign combinations in Table~\ref{tab:sol_cst_eqs} must be considered. Cases (d$_2$) and (g$_2$) have $\frac{d^2\mathcal{T}}{d\delta^2} = 0$ in all regions of $\{\alpha, \kappa, t_x\}$ space; incidentally, such cases have $\frac{d^3\mathcal{T}}{d\delta^3} \neq 0$, thus unambiguously ruling them out as saddle points.
We also find in Section~\ref{subs:bounds} that cases (a), (b), (c), (e), (f) and (h) cannot yield finite time-optimal sequences.  Hence, we establish that internal times in a finite time-optimal $n$-sequence, $n \geq 4$, must satisfy the time relationship described by cases (d$_1$) and (g$_1$),
\begin{equation}
\label{tv:posneg}
\tan\left(\frac{t_v}{2}\right) = \pm \tan\left(\frac{t_x}{2}\right)  \frac{1}{\kappa} \ ,
\end{equation}
with the only possible sign structures being
\begin{equation}
\{+,+,-,-\}, \{-,-,+,+\}, \{+,-,-,+\} \ \textrm{and} \ \{-,+,+,-\} \ .
\end{equation}

In infinite sequences, case (b) is ruled out since
\begin{equation}
\label{caseb}
\lim_{k \to \infty} \left[\mathsf{X}\left(\frac{t_x}{k}\right)\cdot \mathsf{V}\left(\frac{t_v}{k}\right)\cdot\mathsf{X}\left(\frac{t_x}{k}\right)\cdot\mathsf{V}\left(\frac{-t_v}{k}\right)\right]^k = \left[\mathsf{X}(t_x+t_x)\cdot\mathsf{V}(t_v-t_v)\right] = \mathsf{X}(2t_x) \ ;
\end{equation}
in a similar fashion, we rule out cases (c), (e) and (h), which  yield, respectively, $\mathsf{V}(2t_v)$, $\mathsf{X}(2t_x)$ and $\mathsf{V}(2t_v)$. Analogously, cases (d) and (g) are ruled out because
\begin{equation}
\lim_{k \to \infty} \left[\mathsf{X}\left(\frac{t_x}{k}\right)\cdot\mathsf{V}\left(\frac{t_v}{k}\right)\cdot\mathsf{X}\left(\frac{-t_x}{k}\right)\cdot \mathsf{V}\left(\frac{-t_v}{k}\right)\right]^k = \left[\mathsf{X}(t_x-t_x)\cdot \mathsf{V}(t_v-t_v)\right] = \mathds{1} \ .
\end{equation}

We thus establish that time and sign relationships allowed for infinite time-optimal sequences are those described by cases (a) and (f) in Table~\ref{tab:sol_cst_eqs}, namely
\begin{align}
\label{i1}
\tan\left(\frac{t_v}{2}\right) &= \tan\left(\frac{t_x}{2}\right)  K_1 \ , \  \textrm{with signs} \ \{+,+,+,+\},\ \{-,-,-,-\}; \\
\label{i2}
\tan\left(\frac{t_v}{2}\right) &= -\tan\left(\frac{t_x}{2}\right)  \frac{\kappa + \cos(\alpha)}{1+\kappa\cos(\alpha)} \equiv - \tan\left(\frac{t_x}{2}\right)  K_3 \ , \\ 
&\qquad\qquad\qquad \hspace{0.73cm} \textrm{with signs} \ \{+,-,+,-\},\ \{-,+,-,+\}.\nonumber
\end{align}

We already considered case (a), which gives rise to potential solutions via the operator $\mathsf{Q}(t_Q)$ if $\kappa > \cos(\alpha)$; to take into account counter-clockwise rotations, we redefine $t_q$ so that $t_q \in \ ]-\pi, \pi] \ $. 

Case (f) defines a rotation $\mathsf{P}$,
\begin{equation}
\mathsf{P}(t_P) \equiv e^{-i\frac{t_P}{2} (\sigma_x - K_3 \sigma_v)} = \lim_{k \to \infty} \left[\mathsf{X}\left(\frac{t_x}{k}\right)\cdot \mathsf{V}\left(\frac{-t_v}{k}\right)\right]^k  = \lim_{k \to \infty} \left[\mathsf{V}\left(\frac{-t_v}{k}\right)\cdot \mathsf{X}\left(\frac{t_x}{k}\right)\right]^k\ .
\end{equation} 
The normalization of the axis $\vec{n}_p$ is given by 
\begin{equation}
N_p \equiv \frac{\sqrt{1+\kappa^2+2\kappa\cos(\alpha)}\sin(\alpha)}{1+\kappa\cos(\alpha)} \ .
\end{equation}
As previously,
\begin{equation}
\lim_{t_x \to 0} t_v = -t_x \cdot K_3 + \ \smallO{t_x} \ ;
\end{equation}
in order to maintain the alternating sign structure, $K_3 > 0$, which is obtained if and only if $\kappa > \cos(\pi-\alpha)$, defining the regions where a time-optimal sequence involving $\mathsf{P}$ may exist. The time cost associated with a $\mathsf{P}(t_P)$ rotation is $t_P \left(1 + \kappa \cdot K_3 \right)$.  
As for the rotation $\mathsf{Q}$, $t_P \in \ \left]-{\pi}/{N_p}, {\pi}/{N_p} \right] \ $ is again unbounded. For simplicity, we renormalize $t_p \equiv t_P \cdot N_p$, $t_p \in  \ ]-\pi, \pi] \ $.


\subsection{Bounds on internal rotation angles and on maximal length $n$}
\label{subs:bounds}

In the preceding Subsection, we have shown that time-optimal
sequences only depend on three angles. This still leaves undetermined the values of these angles, as
well as the total length of the time-optimal sequence. In what follows, we will
derive bounds for both the sequence length $n$ and the values of the
angles. This not only allows further restricting of the parameter space explored by a numerical search, but also sets constraints on the total time required to synthesize arbitrary unitaries.  

In this Subsection, we concentrate on the sequence length $n$ and on maximal values for the internal angle $t_{x}$ (equivalently, for $t_{v}$). 
For given values of the angle $\alpha$ between rotation axes and the relative rotation speed $\kappa$, we will show that only some values
of $t_{x}$ can occur in time-optimal sequences, a constraint expressed in terms of admissible regions in the $\{t_{x},\alpha,\kappa\}$ space. 
Our core results are obtained by noting that subsequences $U^{\star}$ can have alternative decompositions with different total synthesis times in distinct regions of the parameter space; for given decompositions that satisfy the constraints of Subsection~\ref{subs:relationship}, and that are thus possibly time-optimal, we are often able to find alternative decompositions with a lower synthesis time. This general procedure allows to rule out some decompositions as non-optimal, leading to the definition of the admissible regions.

We extensively use analytical decompositions of a given $U^\star$ into consecutive rotations $\mathsf{A}, \mathsf{B}, \mathsf{C}$ around three non-orthogonal axes $\vec{n}_a, \vec{n}_b, \vec{n}_c$~\cite{Piovan12}. Here we shall choose $\mathsf{A},\mathsf{B},\mathsf{C}$ in the set $\{\mathsf{X},\mathsf{V},\mathsf{Q},\mathsf{P}\}$ so as to obtain alternative decompositions of a given $U^{\star}$ in terms of our control set. We henceforth note this method as \textsf{decomposition} \#1 :
\begin{equation}
U^\star = \mathsf{C}(\theta_3)\cdot \mathsf{B}(\theta_2) \cdot \mathsf{A}(\theta_1) \ .
\end{equation} 
Such decompositions exist if and only if~\cite{Piovan12}
\begin{equation}
|\vec{n}_c^T(u_{\scriptsize{\textrm{goal}}} - \vec{n}_b\vec{n}_b^T)\vec{n}_a| \leq \sqrt{1-(\vec{n}_c^T\vec{n}_b)^2}\sqrt{1-(\vec{n}_a^T\vec{n}_b)^2} \ ,
\end{equation}
where $u_{\scriptsize{\textrm{goal}}}$ is the $SO(3)$ representation of $U_{\scriptsize{\textrm{goal}}} \in SU(2)$ up to a global phase~\cite{RoboticsBook}.
%
When they exist, the decompositions form either a distinct or degenerate pair, with rotation angles $\theta_i \in ]-\pi,\pi]$ given by~\cite{Piovan12}
\begin{align}
\theta_2  &= \textrm{arctan}_2(b,a) \pm \textrm{arctan}_2(\sqrt{a^2+b^2-c^2},c) \ ; \\
\theta_1  &=  -\textrm{arctan}_2(w_a^T \vec{n}_a \times v_a, v_a^T w_a - (v_a^T \vec{n}_a)\cdot (w_a^T \vec{n}_a)) \ ; \\
\theta_3  &=  \textrm{arctan}_2(w_c^T \vec{n}_c \times v_c, v_c^T w_c - (v_c^T \vec{n}_c)\cdot (w_c^T \vec{n}_c)) \ , 
\end{align}
with the definitions
\begin{equation}\begin{array}{ll}
a &= - \vec{n}_c^T \cdot (\textrm{rod}(\vec{n}_b))^2 \cdot \vec{n}_a \ ; \\
b &= \vec{n}_c^T \cdot (\textrm{rod}(\vec{n}_b)) \cdot \vec{n}_a \ ; \\
c &= \vec{n}_c^T \cdot (u_{\scriptsize{\textrm{goal}}} - \mathds{1} - (\textrm{rod}(\vec{n}_b))^2) \cdot \vec{n}_a \ ; \\
v_a &= e^{-\theta_2 \cdot (\textrm{rod}(\vec{n}_b))} \cdot \vec{n}_c \ ; \\
w_a &= u_{\scriptsize{\textrm{goal}}}^T \cdot \vec{n}_c \ ; \\
v_c &= e^{\theta_2 \cdot (\textrm{rod}(\vec{n}_b))} \cdot \vec{n}_a \ ; \\
w_c &= u_{\scriptsize{\textrm{goal}}} \cdot \vec{n}_a \ . \\
\end{array}
\end{equation}
Above, $\textrm{arctan}_2(y, x) =  \textrm{Arg}(x+iy)$  and $\textrm{rod}(\left\{x,y,z\right\})$ is the matrix in Rodrigues' rotation formula~\cite{RoboticsBook}:
\begin{align*}
\textrm{rod}(\{x,y,z\}) &= \left( \begin{array}{ccc}
0 & -z & y \\
z & 0 & -x \\
-y & x & 0 \end{array} \right).
\end{align*}

A special case of the method above is obtained by noting that any 3-subsequence
\begin{equation}
U^\star = \mathsf{A}(\delta)\cdot \mathsf{B}(t)\cdot \mathsf{A}(\delta) \ ,
\end{equation}
$\forall \ |\delta| < |t|$, can be alternatively synthesized as 
\begin{equation}
U^\star = \mathsf{B}(\tau)\cdot \mathsf{A}(\mu)\cdot \mathsf{B}(\tau) \ 
\end{equation}
(\textsf{decomposition} \#2). To first order in $\delta$ the times are
\begin{eqnarray}
\label{tau}
\tau &=& \frac{t}{2} + \delta \ (\vec{n}_a \cdot \vec{n}_b) \left(1-\cos\left(\frac{t}{2}\right)\right) + \ \smallO{\delta^2} \ ; \\
\label{mu}
\mu &=& 2 \ \delta\cos\left(\frac{t}{2}\right) + \ \smallO{\delta^2} \ .
\end{eqnarray}
In the time-optimal synthesis problem, we will consider sequences $\mathsf{V}(t_v)\cdot \mathsf{X}(t_x)\cdot \mathsf{V}(t_v)$ or $\mathsf{X}(t_x)\cdot \mathsf{V}(t_v)\cdot \mathsf{X}(t_x)$. Upon rewriting such sequences as
\begin{align}
\mathsf{V}(t_v)\cdot \mathsf{X}(t_x)\cdot \mathsf{V}(t_v) &= \mathsf{V}(t_v-\delta)\cdot \mathsf{V}(\delta)\cdot \mathsf{X}(t_x)\cdot \mathsf{V}(\delta)\cdot \mathsf{V}(t_v-\delta) \ ; \\
\mathsf{X}(t_x)\cdot \mathsf{V}(t_v)\cdot \mathsf{X}(t_v) &= \mathsf{X}(t_x-\delta)\cdot \mathsf{X}(\delta)\cdot \mathsf{V}(t_v)\cdot \mathsf{X}(\delta)\cdot \mathsf{X}(t_x-\delta) \ ,
\end{align}
with very small $\delta$, we can then apply \textsf{decomposition} \#2 above and rewrite the sequences as
\begin{align}
\mathsf{V}(t_v)\cdot \mathsf{X}(t_x)\cdot \mathsf{V}(t_v) &= \mathsf{V}(t_v-\delta)\cdot \mathsf{X}(\tau)\cdot \mathsf{V}(\mu)\cdot \mathsf{X}(\tau)\cdot \mathsf{V}(t_v-\delta) \ ; \\
\mathsf{X}(t_x)\cdot \mathsf{V}(t_v)\cdot \mathsf{X}(t_x) &= \mathsf{X}(t_x-\delta)\cdot\mathsf{V}(\tau)\cdot \mathsf{X}(\mu)\cdot \mathsf{V}(\tau)\cdot \mathsf{X}(t_x-\delta) \ ,
\end{align}
with $\tau$ and $\mu$ given by Eqs.~\ref{tau},~\ref{mu}, and $(\vec{n}_a \cdot \vec{n}_b) = (\vec{n}_x \cdot \vec{n}_v) = \cos(\alpha)$. In regions of $\{\alpha,\kappa,t_{x}\}$ space where $2\kappa|\delta|+|t_x| > 2|\tau|+\kappa|\mu|$
(respectively, in regions of $\{\alpha,\kappa,t_{v}\}$ space where $2|\delta|+\kappa|t_v|>2\kappa|\tau|+|\mu|$),
the original 3-subsequence $\mathsf{V}(t_v)\cdot \mathsf{X}(t_x)\cdot \mathsf{V}(t_v)$ (respectively, $\mathsf{X}(t_x)\cdot \mathsf{V}(t_v)\cdot \mathsf{X}(t_x)$) synthesizing $U^{\star}$ cannot be time-optimal~\cite{Billig13}. The same method can be applied to infinite sequences of type $\mathsf{X}(\delta)\cdot \mathsf{Q}(t_q)\cdot \mathsf{X}(\delta)$, $\mathsf{V}(\delta)\cdot \mathsf{Q}(t_q)\cdot \mathsf{V}(\delta)$, $\mathsf{X}(\delta)\cdot \mathsf{P}(t_p)\cdot \mathsf{X}(\delta)$ and $\mathsf{V}(\delta)\cdot \mathsf{P}(t_p)\cdot \mathsf{V}(\delta)$. 

Finally, we explore the symmetries that arise when considering rotations such as $[\mathsf{X}(t_x)\mathsf{V}(t_v)] \equiv \mathsf{M}(\theta)$.  $\mathsf{M}(\theta)$ is the effective rotation accomplished by the alternating controls; it is described by an axis $\vec{n}_m \equiv (m_x,m_y,m_z)$, with 

\begin{align}
m_x(t_x,t_v,\alpha) &= \frac{\cos (\alpha) \sin \left(\frac{t_v}{2}\right) \cos \left(\frac{t_x}{2}\right)+\cos \left(\frac{t_v}{2}\right) \sin
   \left(\frac{t_x}{2}\right)}{\sqrt{1-\left(\cos \left(\frac{t_v}{2}\right) \cos \left(\frac{t_x}{2}\right)-\cos (\alpha) \sin \left(\frac{t_v}{2}\right) \sin
   \left(\frac{t_x}{2}\right)\right)^2}} \ ; \\
   m_y(t_x,t_v,\alpha) &= \frac{\sin (\alpha) \sin \left(\frac{t_v}{2}\right) \cos \left(\frac{t_x}{2}\right)}{\sqrt{1-\left(\cos \left(\frac{t_v}{2}\right) \cos
   \left(\frac{t_x}{2}\right)-\cos (\alpha) \sin \left(\frac{t_v}{2}\right) \sin \left(\frac{t_x}{2}\right)\right)^2}} \ ; \\
   m_z(t_x,t_v,\alpha) &= \frac{\sin (\alpha) \sin \left(\frac{t_v}{2}\right) \sin \left(\frac{t_x}{2}\right)}{\sqrt{1-\left(\cos \left(\frac{t_v}{2}\right) \cos
   \left(\frac{t_x}{2}\right)-\cos (\alpha) \sin \left(\frac{t_v}{2}\right) \sin \left(\frac{t_x}{2}\right)\right)^2}} \ ; 
   \end{align} 
   and by the angle 
   \begin{align} 
   \theta(t_x,t_v,\alpha) &= 2 \arccos\left(\cos \left(\frac{t_v}{2}\right) \cos \left(\frac{t_x}{2}\right)-\cos (\alpha ) \sin \left(\frac{t_v}{2}\right) \sin
   \left(\frac{t_x}{2}\right)\right) \ .
\end{align}

We point out that, if $[\mathsf{X}(t_x)\cdot\mathsf{V}(t_v)]$ has axis $(m_x,m_y,m_z)$ and rotation angle $\theta$, related rotations such as $[\mathsf{V}(t_v)\cdot\mathsf{X}(t_x)]$ are similarly parametrized (see Table~\ref{tab:sym_rots}). These relationships allow us to analytically derive alternative decompositions to $U^\star$ composed of three or more consecutive rotations  (\textsf{decomposition} \#3). 

\begin{table}[H]
\centering
\begin{tabular}{|c|c|c|}
\hline
rotation&axis&angle \\
\hline 
$[\mathsf{X}(t_x)\cdot\mathsf{V}(t_v)]$&$(m_x,m_y,m_z)$&$\theta$\\
\hline 
$[\mathsf{X}(-t_x)\cdot\mathsf{V}(-t_v)]$&$(m_x,m_y,-m_z)$&$-\theta$\\
\hline
$[\mathsf{V}(t_v)\cdot\mathsf{X}(t_x)]$ &$(m_x,m_y,-m_z)$&$\theta$\\
\hline
$[\mathsf{V}(-t_v)\cdot\mathsf{X}(-t_x)]$&$(m_x,m_y,m_z)$&$-\theta$\\
\hline
\end{tabular}
\caption{\label{tab:sym_rots} Relationships between parametrization of related rotations.}
\end{table}
The simplest example of these alternative decompositions (\textsf{decomposition} \#4) is obtained by considering that any rotation
\begin{equation}
U^\star =  \mathsf{A}(t) \ 
\end{equation}
can be alternatively synthesized up to a global phase as 
\begin{equation}
U^\star = \mathsf{B}(t^\ast)\cdot \mathsf{A}(-t)\cdot \mathsf{B}(t^\ast) \ ,
\end{equation}
with 
\begin{equation}
t^\ast = - 2\arccot\left( (\vec{n}_a \cdot \vec{n}_b) \ \tan\left(\frac{t}{2}\right) \right) \ .
\end{equation}

In what follows, necessary bounds on the internal rotation angles and on the maximal length $n$ of time-optimal sequences are presented; they are directly derived by fully analytic procedures adopting the
four \textsf{decompositions} described above.  

\vspace{11pt}
\footnotesize $\blacksquare$\normalsize\ \ \underline{\textbf{Case $t >0,\ \kappa > \cos(\alpha)$, finite sequences.}} \ Applied to this case, \textsf{decomposition} \#2 implies that 3-sequences or subsequences of type
\begin{equation}
U^\star = \mathsf{V}(t_f) \cdot \mathsf{X}(t_x) \cdot \mathsf{V}(t_i)
\end{equation}
are only time-optimal for $t_x > \pi$.  For sequences longer than $n = 3$, $t_x>\pi$ implies that $t_v > \pi$ as well. Similarly, 3-sequences such as
\begin{equation}
U^\star = \mathsf{X}(t_f) \cdot \mathsf{V}(t_v) \cdot \mathsf{X}(t_i)
\end{equation}
are only time-optimal for $t_v > \pi$; equivalently, for sequences with $n \geq 3$, $t_x > \pi$ implies $t_v > \pi$ as well.

To further bound the allowed $t_x$ and sequence length $n$, we focus on the case $\kappa=1$ and show that for some ranges of $\{\alpha, t_x\}$, time-optimal sequences with finite length greater or equal to a given $n$ do not exist. 
While for simplicity we omit details for the case $\kappa \neq 1$, we note that  our methods can be extended in a straightforward way to rotations with different implementation speeds. In addition, we observe that, given a sequence of length $n$, the allowed regions for time-optimal sequences in $\{\alpha,  t_x\}$ space expand with increasing $\kappa$.
Thus, although a formal proof is lacking,  the limit $\kappa = 1$ may be taken as a loose bound for the necessary structure of a time-optimal sequence. 

\begin{figure}[t]
\centering
\includegraphics[width=0.34\columnwidth]{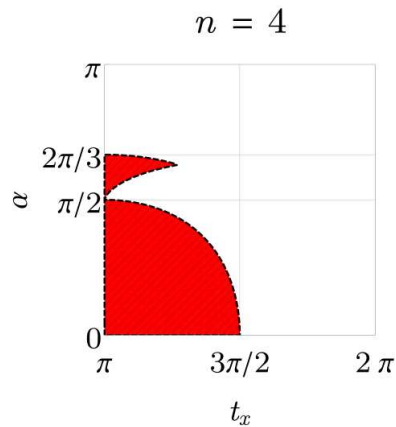}
\caption{Regions in $\{\alpha, t_x\}$ space where a $n = 4$ sequence can be time-optimal, in the case $\kappa = 1$, are depicted in dashed red. The region for which $\alpha < \frac{\pi}{2}$ is described by $t_x \leq t^\dagger$, with $t^\dagger$ defined as the angle for which $[\mathsf{X}(t^\dagger) \cdot \mathsf{V}(t^\dagger)] = [\mathsf{V}(-t^\dagger) \cdot \mathsf{X}(-t^\dagger)]$.}
\label{fig:4}
\end{figure}

A 4-sequence can only be time-optimal in the regions shown in Figure~\ref{fig:4}. There are several ways of deriving this result; one of them is to overlay the regions in $\{\alpha, t_x\}$ space where, concomitantly, $\frac{d^2\mathcal{T}}{d\delta^2} > 0$ and one alternative decomposition of $[\mathsf{X}(t_x)\cdot\mathsf{V}(t_v)]$, for example
\begin{align}
[\mathsf{X}(t_x)\cdot\mathsf{V}(t_v)] &= \mathsf{V}(\theta_3)\cdot  \mathsf{Q}(\theta_2)\cdot \mathsf{X}(\theta_1) \ \textrm{(\textsf{decomposition} \#1)} \ ; \\
[\mathsf{X}(t_x)\cdot\mathsf{V}(t_v)] &= \mathsf{V}(\theta_3)\cdot  \mathsf{X}(\theta_2)\cdot \mathsf{V}(\theta_1)\cdot \mathsf{X}(-t_x) \ \textrm{(\textsf{decompositions} \#1,3)} \ ,
\end{align} 
is synthesized in less time. Here and in the following, negative rotation angles such as $-t_x$ should be interpreted as implemented by physical rotations by the positive angle $2\pi - t_x$. There are two distinct regions\footnote{The appearance of the region with $\pi/2<\alpha<2\pi/3$ 
is not intuitive; since the marked zones only reflect necessary conditions for time-optimality, we independently confirm the existence of the two disjoint regions with numerical simulations.} in Figure~\ref{fig:4}; the region for $\alpha < \frac{\pi}{2}$ is given by $t_x \leq t^\dagger$, where $t^\dagger$ is the angle for which $[\mathsf{X}(t^\dagger) \cdot \mathsf{V}(t^\dagger)] = [\mathsf{V}(-t^\dagger) \cdot \mathsf{X}(-t^\dagger)]$; the significance of angles of high sequence symmetry such as $t^\dagger$ will be further explored below.
For orthogonal controls $\alpha = \frac{\pi}{2}$, we remark that finite $n \geq 4$ sequences are never time-optimal; to our knowledge, this is an original proof  that time-optimal sequences using orthogonal controls are achieved either with 3-long Euler-like decompositions, or with an infinite concatenation of controls. 

Finally, note that the second derivative argument of Subsection~\ref{subs:relationship} had already ruled out $n \geq 4$ or longer finite subsequences for $\alpha > \frac{2\pi}{3}$ as non-optimal. 

For longer sequences with $n \geq 5$, two alternative decompositions can be employed based on \textsf{decompositions} \#1,3, namely
\begin{align}
[\mathsf{X}(t_x)\cdot\mathsf{V}(t_v)]^k \cdot \mathsf{X}(t_x) &= \mathsf{X}(\theta_3) \cdot [\mathsf{X}(-t_x)\cdot\mathsf{V}(-t_v) ]^k \cdot \mathsf{X}(-t_x) \cdot \mathsf{V}(\theta_1)\ , \label{eq:odd2}\\
[\mathsf{X}(t_x)\cdot\mathsf{V}(t_v)]^k \cdot \mathsf{X}(t_x) &= \mathsf{X}(\theta_3) \cdot [\mathsf{X}(-t_x)\cdot\mathsf{V}(-t_v) ]^k \cdot \mathsf{X}(-t_x) \cdot \mathsf{X}(\theta_1)\ ; \label{eq:odd1}
\end{align}
if $n$ is odd; and, for even $n$,
\begin{align}
[\mathsf{X}(t_x)\cdot\mathsf{V}(t_v)]^k &= \mathsf{X}(\theta_3) \cdot [\mathsf{V}(-t_v)\cdot\mathsf{X}(-t_x) ]^k \cdot \mathsf{X}(\theta_1) \ ; \label{eq:even1}\\
[\mathsf{X}(t_x)\cdot\mathsf{V}(t_v)]^k &= \mathsf{X}(\theta_3) \cdot [\mathsf{X}(-t_x)\cdot\mathsf{V}(-t_v) ]^k \cdot \mathsf{V}(\theta_1) \ .\label{eq:even2}
\end{align}
Using these decompositions, for $\kappa = 1$, we obtain a consistent scaling law for the regions in $\{\alpha, t_x\}$ space where time-optimal sequences of length $n \geq 5$ can exist. 

\begin{figure}[t]
\centering
\includegraphics[width=0.79\columnwidth]{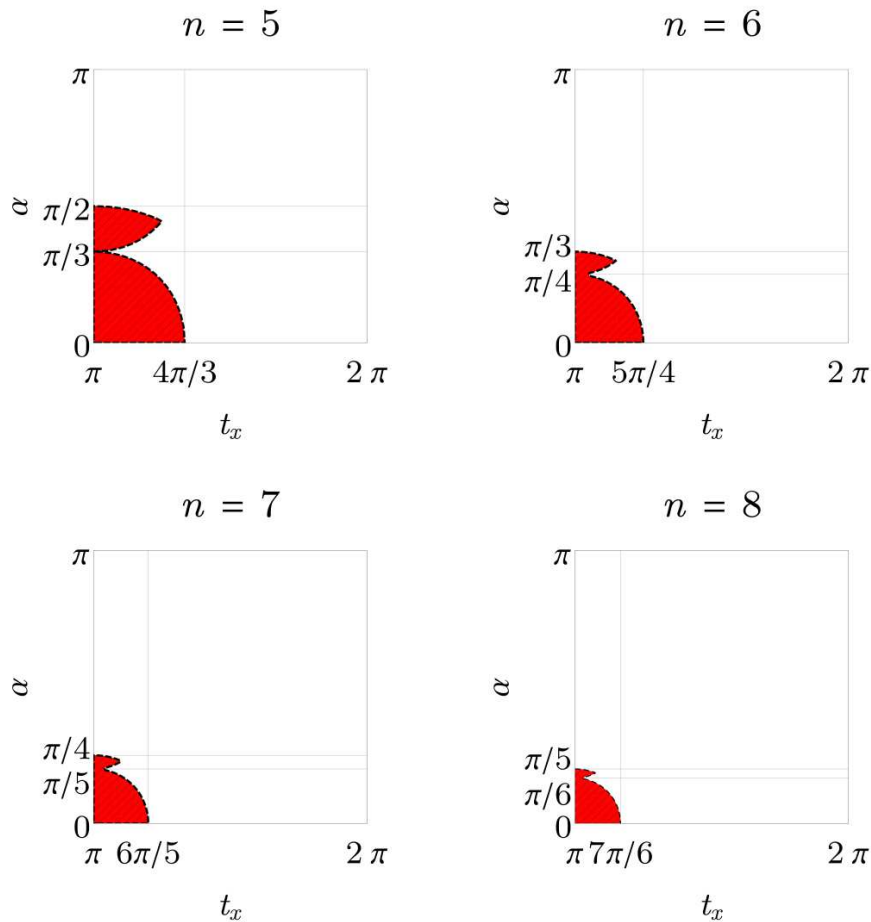}
\caption{Regions in $\{\alpha, t_x\}$ where time-optimal sequences of length $n \geq 5$ can exist, for the particular case $\kappa = 1$, are depicted in dashed red. Note the scaling laws $t_x \leq \frac{(n-1)}{(n-2)}\pi$, and $\alpha \leq \frac{1}{(n-3)}\pi$. }
\label{fig:5678}
\end{figure}

Define the rotation angles $t_{\scriptsize{\textrm{odd,$k$}}},t_{\scriptsize{\textrm{even,$k$}}}$ such that
\begin{align}
[\mathsf{X}(t_{\scriptsize{\textrm{odd,$k$}}})\cdot\mathsf{V}(t_{\scriptsize{\textrm{odd,$k$}}})]^k \cdot\mathsf{X}(t_{\scriptsize{\textrm{odd,$k$}}}) &= [\mathsf{V}(-t_{\scriptsize{\textrm{odd,$k$}}})\cdot\mathsf{X}(-t_{\scriptsize{\textrm{odd,$k$}}})]^k \cdot\mathsf{V}(-t_{\scriptsize{\textrm{odd,$k$}}}) \ ; \\
[\mathsf{X}(t_{\scriptsize{\textrm{even,$k$}}})\cdot\mathsf{V}(t_{\scriptsize{\textrm{even,$k$}}})]^k &= [\mathsf{V}(-t_{\scriptsize{\textrm{even,$k$}}})\cdot\mathsf{X}(-t_{\scriptsize{\textrm{even,$k$}}})]^k \ .
\end{align} 
Such angles are explicitly given by
\begin{align}
t_{\scriptsize{\textrm{odd,$k$}}} &= 2\arccos\left(-\sqrt{\frac{\cos(\alpha) - \cos\left(\frac{\pi}{2k + 1} \right)}{1+\cos(\alpha)}} \right) \ ; \\
t_{\scriptsize{\textrm{even,$k$}}} &= 2\arccos\left(-\sqrt{\frac{\cos(\alpha) - \cos\left(\frac{\pi}{2k} \right)}{1+\cos(\alpha)}} \right) \ .
\end{align}
Now, for an odd $n = (2k + 3)$, $n \geq 5$, time-optimal $n$-sequences with middle rotation angle $t_x$ can exist for $t_x \leq t_{\scriptsize{\textrm{odd,$k$}}}$ and for a small region such that $t_{\scriptsize{\textrm{odd,$k$}}} \leq t_x \leq t_{\scriptsize{\textrm{even,$k$}}}$. These relationships are obtained by employing, respectively,~\req{eq:odd2} and~\req{eq:odd1}. 
Similarly, for an even $n = (2k + 2)$, $n \geq 6$, time-optimal $n$-sequences must have the middle rotation angle  $t_x \leq t_{\scriptsize{\textrm{even,$k$}}}$ or within a small region given by $t_{\scriptsize{\textrm{even,$k$}}} \leq t_x \leq t_{\scriptsize{\textrm{odd,$(k-1)$}}}$, as obtained from equations \req{eq:even1} and~\req{eq:even2}, respectively. This situation is depicted in Figure~\ref{fig:5678}.  

\begin{figure}
\centering
\includegraphics[width=0.825\columnwidth]{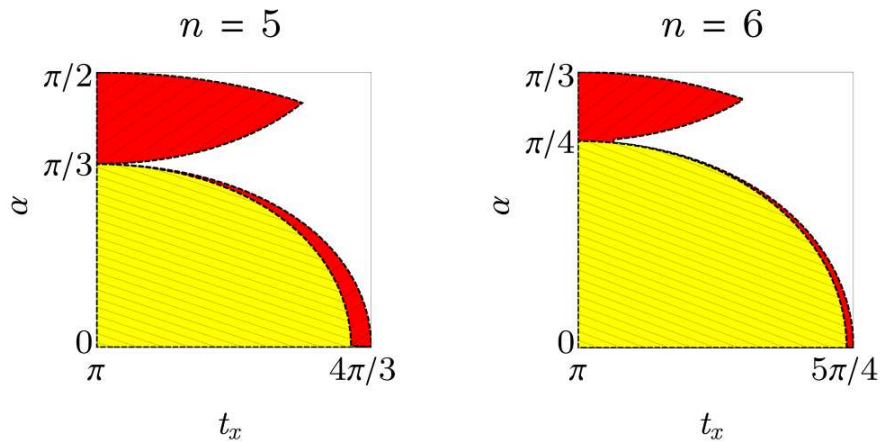}
\caption{Stricter regions in $\{\alpha, t_x\}$ space where a $n = 5,6$ sequence can be time-optimal, in the case $\kappa = 1$, are depicted in fine-dashed yellow; bounds obtained in Figure~\ref{fig:5678} are in dashed red.}
\label{fig:56new}
\end{figure}

For the particular cases $n = 5, 6$, we find, however, tighter bounds using the following alternative decompositions:
\begin{align}
\mathsf{X}(t_x)\cdot\mathsf{V}(t_v) \cdot \mathsf{X}(t_x) &= \mathsf{V}(\theta_3) \cdot \mathsf{X}(\theta_2) \cdot \mathsf{V}(\theta_1) \   \textrm{(\textsf{decomposition} \#1)} \ ;\\
[\mathsf{X}(t_x)\cdot\mathsf{V}(t_v)]^2 &= \mathsf{V}(\theta_3) \cdot \mathsf{X}(\theta_2) \cdot \mathsf{V}(\theta_1) \cdot \mathsf{X}(-t_x) \ \textrm{(\textsf{decompositions} \#1,3)} \ .
\end{align}
These new decompositions completely cut the small disjoint region at higher $\alpha$ (while further constraining the maximal $t_x < t_{\scriptsize{\textrm{even,$k$}}}, t_{\scriptsize{\textrm{odd,$k$}}}$). The viable smaller regions are plotted in Figure~\ref{fig:56new} against the previous bounds shown in Figure~\ref{fig:5678}. Numerical simulations confirm these tighter bounds. Thus, we conjecture that there might be other decomposition of $n \geq 7$ sequences that remove the disjoint region for those longer sequences as well, although this does not appear to be the case for $n=4$.

To sum up, $n = 4$ time-optimal sequences are bounded by $t_x < \frac{3\pi}{2}$ and $\alpha < \frac{2\pi}{3}$, while for $n \geq 5$ they satisfy $t_x \leq \frac{(n-1)}{(n-2)}\pi$, and $\alpha \leq \frac{\pi}{n-3}$ (with a plausible tighter limit at $\alpha \leq \frac{\pi}{n-2}$). 

Inverting the constraints on the admissible regions, $\alpha(n) \rightarrow n(\alpha)$, we find new bounds on the maximum length of a time-optimal sequence: 
\begin{align}
n \leq \lfloor\frac{\pi}{\alpha}\rfloor + 3, \ \mathrm{for }\ n \geq 5 \ .
\end{align}
Note that, especially for small $\alpha \leq \frac{\pi}{3}$, this is a much tighter bound than those previously obtained with index theory~\cite{Agrachev90}, which predicts that a finite time-optimal sequence would bear no more than $n \leq \lfloor\frac{2\pi}{\alpha}\rfloor$ control concatenations; and with geometric control~\cite{Boscain05}, which sets $n \leq \lfloor\frac{\pi}{\alpha}\rfloor + 5$.   

\vspace{11pt}
\footnotesize $\blacksquare$\normalsize\ \ \underline{\textbf{Case $t >0, \kappa > \cos(\alpha)$, infinite sequences.}} \ Infinite sequences that are time-optimal must necessarily be of form
$\mathsf{X}(t_f) \cdot \mathsf{Q}(t_Q)\cdot  \mathsf{V}(t_i)$ or $\mathsf{V} (t_f)\cdot\mathsf{Q}(t_Q)\cdot  \mathsf{X}(t_i)$, with $t_q < \pi$. 
This result stems from the fact that \textsf{decomposition} \#2 imposes that an infinite sequence of the form $\mathsf{A}(t_f)\cdot \mathsf{Q}(t_Q)\cdot \mathsf{A}(t_i)$ can only be optimal for $t_q > \pi$; whereas \textsf{decomposition} \#4 requires that an optimal infinite sequence of any form must have $t_q < \pi$. By contradiction, we conclude that an infinite sequence may only be time-optimal in the forms outlined above. 

\vspace{11pt}
\footnotesize $\blacksquare$\normalsize\ \ \underline{\textbf{Case $t >0,\ \kappa < \cos(\alpha)$.}} \ \textsf{Decomposition} \#2 requires that 3 or longer time-optimal sequences have $t_x < \pi$. Note that $t_v(t_x) > \pi$.  
Although we cannot simply find further bounds  for $n = 4,5$-long time-optimal sequences, a straightforward application of \textsf{decomposition} \#1 constrains $n = 6$ or longer  time-optimal sequences  to have $t_x > \frac{\pi}{3}$. Additionally, using the same {decomposition} we find that, if $\alpha > \textrm{min}\{\frac{\pi}{1+k}, \arccos(\kappa)\}$, an $n = (2k + 2), (2k + 3)$ sequence, with $k \geq 2$, cannot be time-optimal.   

Hence, if $\frac{\pi}{1+k} \leq \arccos(\kappa)$, one can place a bound on the maximal length of a time-optimal sequence:
\begin{equation}
n \leq \lfloor\frac{2\pi}{\alpha}\rfloor + 1, \ \mathrm{for }\ n \geq 6 \ .
\end{equation}
\vspace{11pt}
\footnotesize $\blacksquare$\normalsize\ \ \underline{\textbf{Case $t \lessgtr 0$, finite sequences.}} \ Bounds on the maximal length of a finite time-optimal sequence are readily obtained. In particular, for all $n \geq 6$  finite time-optimal sequences,  there is always at least one alternative decomposition, $\forall \ \kappa$,  that synthesizes the same unitary in a shorter time. For example, considering a 4-subsequence of a $n \geq 6$ sequence, the unitary realized by the inner rotations such as
\begin{equation}
U^\star = \mathsf{X}(+t_x)\cdot [\mathsf{V}(+t_v)\cdot \mathsf{X}(-t_x)]\cdot \mathsf{V}(-t_v)
\end{equation}
has alternative decompositions:
\begin{align}
U^\star &= \mathsf{V}(\theta_3)\cdot \mathsf{X}(\theta_2)\cdot \mathsf{V}(\theta_1) \ \textrm{(\textsf{decomposition} \#1)} \ ;  \\
U^\star &= \mathsf{X}(+t_x)\cdot(\mathsf{V}(\theta_3)\cdot [\mathsf{X}(+t_x)\cdot\mathsf{V}(-t_v)]\cdot \mathsf{V}(\theta_1))\cdot\mathsf{V}(-t_v)  \ \textrm{(\textsf{decompositions} \# 1,3)}\ ; \\
U^\star &= \mathsf{X}(+t_x)\cdot(\mathsf{X}(\theta_3)\cdot [\mathsf{V}(-t_v)\cdot\mathsf{X}(+t_x)]\cdot\mathsf{V}(\theta_1))\cdot \mathsf{V}(-t_v)  \ \textrm{(\textsf{decompositions} \# 1,3)} \ ,
\end{align}
with at least one of the above having a lower total synthesis time, in all regions of $\{\alpha, \kappa, t_x\}$ space. 

We thus proved the important result that, for $t \lessgtr 0$, time-optimal sequences must be $n \leq 5$ long, or infinite. 

This conclusion is stronger than previously published bounds establishing that every time-optimal trajectory
is a finite concatenation of at most five bang-bang or singular arcs~\cite{Boscain02}, in that our results rule out, for example,
the existence of a time-optimal sequence of type $ \mathsf{A}(t_5)\cdot  \mathsf{B}(t_4)\cdot \mathsf{Q}(t_3 )\cdot \mathsf{B}(t_2) \cdot \mathsf{A}(t_1)$. 

For this case, we can further characterize  the admissible time-optimal sequences  and impose stricter constraints on their times.  3-sequences or subsequences of type
\begin{equation}
U^\star = \mathsf{X}(t_f) \cdot \mathsf{V}(t_v) \cdot \mathsf{X}(t_i) 
\end{equation}
must, according to \textsf{decomposition} \#2, have $\sgn(t_f) \neq \sgn(t_i)$ and $|t_v| < 2\arccos\left( \frac{\kappa |\cos(\alpha)|}{1+\kappa |\cos(\alpha)|} \right)$. 3-sequences or subsequences of type
\begin{equation}
U^\star = \mathsf{V}(t_f) \cdot \mathsf{X}(t_x) \cdot \mathsf{V}(t_i) \ ,
\end{equation}
 must, in turn, obey $\sgn(t_f) \neq \sgn(t_i)$ and $|t_x| < 2\arccos\left( \frac{|\cos(\alpha)|}{\kappa+ |\cos(\alpha)|} \right)$. Time-optimal sequences of this type which are exactly $n = 3$ long may also have $\sgn(t_f) = \sgn(t_i)$; if so, $|t_x| < 2\arccos\left( \frac{\cos(\alpha)-\kappa}{\cos(\alpha)+\kappa} \right)$ should hold. 

Applying \textsf{decomposition} \#4 to the two possible inner subsequences of a 5-long time-optimal sequence, namely
\begin{align}
U^\star &= \mathsf{V}(-t_v) \cdot \mathsf{X}(t_x) \cdot \mathsf{V}(t_v) \ ; \\
U^\star &= \mathsf{X}(-t_x) \cdot \mathsf{V}(t_v) \cdot \mathsf{X}(t_x) \ ,
\end{align}
we conclude that $|t_x| < 2\arccot\left( \frac{1}{\kappa}\right)$ in the first case, and $|t_x| < \frac{\pi}{2}$ in the second.  \\

\footnotesize $\blacksquare$\normalsize\ \ \underline{\textbf{Case $t \lessgtr 0$, infinite sequences.}} \ Because of the sign structure imposed on time-optimal sequences for this case, some forms of infinite sequences can be ruled out. 
In particular, it follows from \textsf{decomposition} \#2 that infinite time-optimal sequences can only take one of the following shapes: $\mathsf{X}(t_f) \cdot \mathsf{Q}(t_Q) \cdot \mathsf{V}(t_i)$, $\mathsf{V}(t_f) \cdot \mathsf{Q}(t_Q) \cdot \mathsf{X}(t_i)$, with $\sgn(t_i) = \sgn(t_f) = \sgn(t_Q)$; or $\mathsf{A}(t_f) \cdot \mathsf{P}(t_P) \cdot \mathsf{B}(t_i)$, with $\sgn(t_i) \neq \sgn(t_f)$. In the latter case, if $\mathsf{A}(\cdot) = \mathsf{B}(\cdot) = \mathsf{X}(\cdot)$ (respectively, $\mathsf{A}(\cdot) = \mathsf{B}(\cdot) = \mathsf{V}(\cdot)$), according to \textsf{decomposition} \#4, $|t_p| < \frac{2\pi}{3}$ ($|t_p| < 2\arccos\left(\frac{\kappa}{1+\kappa}\right)$).

\subsection{Bounds on outer rotation angles}
\label{outer}

Using the same methods as those outlined in Subsection~\ref{subs:bounds}, outer rotation angles can also be constrained. We denote those angles $t_{x,\scriptsize{\textrm{out}}}$ ($t_{v,\scriptsize{\textrm{out}}}$) if
\begin{equation}
U_{\scriptsize{\textrm{goal}}} = \mathsf{X}(t_{x,\scriptsize{\textrm{out}}}) \cdot \ ...  \ \ \textrm{or} \ \  U_{\scriptsize{\textrm{goal}}} =  \ ... \ \cdot \mathsf{X}(t_{x,\scriptsize{\textrm{out}}}) \ .
\end{equation}

\vspace{11pt}
\footnotesize $\blacksquare$\normalsize\ \ \underline{\textbf{Case $t >0,\ \kappa > \cos(\alpha)$.}} For any $n \geq 4$ sequence, we find loose bounds for the outer times by employing \textsf{decomposition} \#4. 

For $\alpha > \frac{\pi}{2}$,  we have $t_{x,\scriptsize{\textrm{out}}}, t_{v,\scriptsize{\textrm{out}}} < \pi$; 
for $\alpha < \frac{\pi}{2}$, we obtain $t_{x,\scriptsize{\textrm{out}}} < 2\pi + 2\arccot\left( \cos(\alpha) \tan\left(\frac{t_v}{2}\right) \right)$  and, similarly, $t_{v,\scriptsize{\textrm{out}}} < 2\pi + 2\arccot\left( \cos(\alpha) \tan\left(\frac{t_x}{2}\right) \right)$. Note that these last two  bounds are tighter than the simpler bounds $t_{\{x,v\},\scriptsize{\textrm{out}}}< 3\pi - t_{\{v,x\}}$. 

\vspace{11pt}
\footnotesize $\blacksquare$\normalsize\ \ \underline{\textbf{Case $t \lessgtr 0$.}} When applied to 3-subsequences such as
\begin{equation}
U^\star = \mathsf{V}(-t_{v,\scriptsize{\textrm{out}}}) \cdot \mathsf{X} (t_x) \cdot \mathsf{V} (t_v) \ ,
\end{equation} 
\textsf{decomposition} \#4 dictates that $|t_{\{x,v\},\scriptsize{\textrm{out}}}| < \pi - |t_{\{x,v\}}|$. Similarly, for a 4-sequence to be time-optimal, $|t_{v,\scriptsize{\textrm{out}}}| + |t_{x,\scriptsize{\textrm{out}}}| < \frac{2\pi}{3}$; this bound is further tightened for a 5-sequence, for which either $|t_{v,\scriptsize{\textrm{out}}}| < t_v$, $|t_{x,\scriptsize{\textrm{out}}}| < t_x$ or $|t_{v,\scriptsize{\textrm{out}}}|, |t_{x,\scriptsize{\textrm{out}}}| < \frac{\pi}{3}$ (whichever is tighter).

\subsection{Summary of results}
\label{subs:summary}
We present a summary of the derived necessary conditions for time-optimal sequences of length $3 \leq n \leq \infty$. As with the rest of this work, the results are subdivided by cases. 

\vspace{11pt}
\footnotesize $\blacksquare$\normalsize\ \ \underline{\textbf{Case $t > 0$.}} Time-optimal sequences only depend on four
parameters, namely the outer angles $t_i, t_f$, the internal angle $t_{x}$ (or $t_v$ or $t_q$), and the total number of rotations $n \leq \infty$. In this case, if $n\geq4$, the internal angles are related by \req{eq:tv_pos},
\begin{equation}
\tan\left(\frac{t_v}{2}\right) = \tan\left(\frac{t_x}{2}\right) \cdot \frac{\kappa - \cos(\alpha)}{1 - \kappa \cos(\alpha)} \ . \tag{14}
\end{equation}
Admissible time-optimal sequences and their derived bounds are summarized in Table~\ref{T1} for $\kappa > \cos(\alpha)$, and in Table~\ref{T2} for $\kappa < \cos(\alpha)$. 

We can further provide bounds on the maximum total time $\mathcal{T}_{\textrm{max}}$ required for a finite time-optimal sequence. 

For the case $\kappa > \cos(\alpha)$, noting that
\begin{align}
\mathrm{max}(t_x + t_v) &= 2\pi + \frac{(n-1)}{(n-2)}\pi = \frac{(3n-5)}{(n-2)}\pi \ ; \\
\mathrm{max}(t_{x,\mathrm{out}} + t_v) &= \pi(2+\kappa) \ ; \\
\mathrm{max}(t_{v,\mathrm{out}} + t_x) &= \pi(1+2\kappa) \ ,
\end{align}
we establish the following:
\begin{itemize}
\item For $n$ odd and outer controls $\mathsf{X}$, $\mathcal{T}_{\textrm{max}} = 2\pi(2+\kappa) + \mathds{1}_{n > 3} \cdot \left( \frac{(n-5)}{2}\frac{(3n-5)}{(n-2)}\pi + \frac{(n-1)}{(n-2)}\pi \right)$;
\item For $n$ odd and outer controls $\mathsf{V}$, $\mathcal{T}_{\textrm{max}} = 2\pi(1+2\kappa) + \mathds{1}_{n > 3} \cdot \left( \frac{(n-5)}{2}\frac{(3n-5)}{(n-2)}\pi + 2\pi\kappa \right)$;
\item For $n \geq 4$ even, $\mathcal{T}_{\textrm{max}} = 3\pi(1+\kappa) + \frac{(n-4)}{2}\frac{(3n-5)}{(n-2)}\pi$.
\end{itemize}

Similarly, for $\kappa < \cos(\alpha)$, given
\begin{align}
\mathrm{max}(t_x + t_v) &= 3\pi \ ,
\end{align}
the maximum times follow:
\begin{itemize}
\item For $n$ odd and outer controls $\mathsf{X}$, $\mathcal{T}_{\textrm{max}} = 2\pi(2+\kappa) +  \frac{(n-3)}{2}3\pi $;
\item For $n$ odd and outer controls $\mathsf{V}$, $\mathcal{T}_{\textrm{max}} = \pi(1+4\kappa) +  \frac{(n-3)}{2}3\pi $;
\item For $n \geq 4$ even, $\mathcal{T}_{\textrm{max}} = 2\pi(1+\kappa) + \frac{(n-2)}{2}3\pi$.
\end{itemize}

We note that these are quite loose bounds, since they are obtained by combining bounds on all free parameters; they might still be of guidance when designing practical experiments.

\vspace{11pt}
\footnotesize $\blacksquare$\normalsize\ \ \underline{\textbf{Case $t \lessgtr 0$.}}  Time-optimal sequences only depend on four
parameters, namely the outer angles $t_i, t_f$, the internal angle $t_{x}$ (or $t_v$ or $t_q$), and the total number of rotations. All angles $t \in [-\pi, \pi].$  It holds that either $n \leq 5$, or $n \to \infty$; moreover, the relative signs of the rotation angles are restricted to a few combinations. If $n\geq4$, internal angles are related by \req{tv:posneg},
\begin{equation}
\tan\left(\frac{t_v}{2}\right) = \pm \tan\left(\frac{t_x}{2}\right) \cdot \frac{1}{\kappa}  \tag{18}
\end{equation}
in finite sequences; and, in infinite sequences, by Eqs.~\ref{i1} and \ref{i2},
\begin{equation}
\tan\left(\frac{t_v}{2}\right) =  \tan\left(\frac{t_x}{2}\right) \cdot \frac{\kappa - \cos(\alpha)}{1 - \kappa \cos(\alpha)} \ ; \\ \tag{22}
\end{equation}
\begin{equation}
\hspace{0.25cm} \tan\left(\frac{t_v}{2}\right) = -\tan\left(\frac{t_x}{2}\right) \cdot \frac{\kappa + \cos(\alpha)}{1+\kappa\cos(\alpha)} \ . \tag{23}
\end{equation}
Admissible time-optimal sequences and their derived bounds are summarized in Table~\ref{T3}. These bounds further provide constraints on the total time of an optimal unitary synthesis. Analogously as above, the maximal total time $\mathcal{T}_{\textrm{max}}$ for a finite time-optimal sequence can be estimated:
\begin{itemize}
\item For $n = 3$ and outer controls $\mathsf{X}$, $\mathcal{T}_{\textrm{max}} = \pi(4 + \kappa)$;
\item For $n = 3$ and outer controls $\mathsf{V}$, $\mathcal{T}_{\textrm{max}} = \pi(1 + 4\kappa)$;
\item For $n = 4$, $\mathcal{T}_{\textrm{max}} = \pi(1 + \kappa)$;
\item For $n = 5$ and outer controls $\mathsf{X}$, $\mathcal{T}_{\textrm{max}} = \frac{13\pi}{6} + \pi\kappa$;
\item For $n = 5$ and outer controls $\mathsf{V}$, $\mathcal{T}_{\textrm{max}} = 2\pi + \frac{2\pi}{3}\kappa$.
\end{itemize}

\begin{table}[H]
\centering
\resizebox{\columnwidth}{!}{%
\begin{tabular}{|c|c|l|c|c|}
\hline
$n$&$\alpha$ range&\hspace{3cm} type & internal angle & outer angles \\
\hline 
\multirow{2}{*}{3} & \multirow{2}{*}{$\alpha < \pi$} & $\mathsf{V}(t_f)\cdot \mathsf{X}(t_x) \cdot \mathsf{V}(t_i)$ & $\pi < t_x$& \\
\hhline{~~--~}
 & & $\mathsf{X}(t_f)\cdot \mathsf{V}(t_v) \cdot \mathsf{X}(t_i)$ & $\pi < t_v$& \\
\hhline{-----}
\multirow{2}{*}{4} & \multirow{2}{*}{$\alpha \leq \frac{2\pi}{3}$} & $\mathsf{V}(t_f)\cdot [\mathsf{X}(t_x) \cdot \mathsf{V}(t_v)] \cdot \mathsf{X}(t_i)$ & \multirow{2}{*}{$\pi < t_x < \frac{3\pi}{2}$} & \multirow{6}{*}{$\begin{cases} t_i, t_f < \pi \mbox{ if } \alpha > \frac{\pi}{2} \mbox{ ($n = 4$ only)}\\ t_{x,\scriptsize{\textrm{out}}} < 3\pi - t_v, \\
t_{v,\scriptsize{\textrm{out}}} < 3\pi - t_x  \mbox{ if } \alpha < \frac{\pi}{2} 
\end{cases}$}\\
\hhline{~~-~~}
 & & $\mathsf{X}(t_f)\cdot [\mathsf{V}(t_v) \cdot \mathsf{X}(t_x)] \cdot \mathsf{V}(t_i)$ &  & \\
\hhline{----~}
\multirow{4}{*}{\small{4 $< n \leq \lfloor\frac{\pi}{\alpha}\rfloor + 3$}} &
\multirow{4}{*}{$\alpha \leq \frac{\pi}{n-3}$}   & $\mathsf{V}(t_f)\cdot [\mathsf{X}(t_x) \cdot \mathsf{V}(t_v)]^{k} \cdot \mathsf{X}(t_x) \cdot \mathsf{V}(t_i), \ k \geq 1$ & \multirow{4}{*}{\small{$\pi < t_x \leq \frac{(n-1)}{(n-2)}\pi$}} & \\
\hhline{~~-~~}
 & &  $\mathsf{V}(t_f)\cdot [\mathsf{X}(t_x) \cdot \mathsf{V}(t_v)]^{k} \cdot \mathsf{X}(t_i), \ k \geq 2$ & & \\
\hhline{~~-~~}
 & & $\mathsf{X}(t_f)\cdot [\mathsf{V}(t_v) \cdot \mathsf{X}(t_x)]^{k} \cdot \mathsf{V}(t_v) \cdot \mathsf{X}(t_i), \ k \geq 1$  & & \\
\hhline{~~-~~}
 & & $\mathsf{X}(t_f)\cdot [\mathsf{V}(t_v) \cdot \mathsf{X}(t_x)]^{k} \cdot \mathsf{V}(t_i), \ k \geq 2$ & & \\
\hline
\multirow{2}{*}{$\infty$} & \multirow{2}{*}{\small{$\arccos(\kappa) < \alpha < \pi$}} & $\mathsf{V}(t_f)\cdot \mathsf{Q}(t_Q) \cdot \mathsf{X}(t_i)$ & \multirow{2}{*}{$t_q < \pi$} & \\
\hhline{~~-~~}
& & $\mathsf{X}(t_f)\cdot \mathsf{Q}(t_Q) \cdot \mathsf{V}(t_i)$ & & \\
\hline
\end{tabular}
}
\caption[Admissible structures of time-optimal sequences, case $t > 0, \ \kappa > \cos(\alpha)$]{\label{T1}Admissible structures of time-optimal sequences, case $t > 0, \ \kappa > \cos(\alpha)$.}
\end{table}

\begin{table}[H]
\centering
\resizebox{0.8\columnwidth}{!}{%
\begin{tabular}{|c|c|l|c|}
\hline
$n$&$\alpha$ range&\hspace{3cm} type & internal angle\\
\hline 
\multirow{2}{*}{3} & \multirow{6}{*}{$0 < \alpha < \arccos(\kappa)$} & $\mathsf{V}(t_f)\cdot \mathsf{X}(t_x) \cdot \mathsf{V}(t_i)$ & $t_x < \pi$\\
\hhline{~~--}
 & & $\mathsf{X}(t_f)\cdot \mathsf{V}(t_v) \cdot \mathsf{X}(t_i)$ & $\pi < t_v$ \\
\hhline{-~--}
\multirow{2}{*}{4} &  & $\mathsf{V}(t_f)\cdot [\mathsf{X}(t_x) \cdot \mathsf{V}(t_v)] \cdot \mathsf{X}(t_i)$ & \multirow{4}{*}{$t_x < \pi$}\\
\hhline{~~-~}
 & & $\mathsf{X}(t_f)\cdot [\mathsf{V}(t_v) \cdot \mathsf{X}(t_x)] \cdot \mathsf{V}(t_i)$ &  \\
 \hhline{-~-~}
\multirow{2}{*}{5} &  & $\mathsf{V}(t_f)\cdot [\mathsf{X}(t_x) \cdot \mathsf{V}(t_v)] \cdot \mathsf{X}(t_x) \cdot \mathsf{V}(t_i)$ & \\
\hhline{~~-~}
 & & $\mathsf{X}(t_f)\cdot [\mathsf{V}(t_v) \cdot \mathsf{X}(t_x)] \cdot \mathsf{V}(t_v) \cdot \mathsf{X}(t_i)$ &  \\
\hhline{----}
\multirow{4}{*}{\small{6 $\leq n < \infty$}} &
\multirow{4}{*}{$\alpha \leq \textrm{min}\{\frac{\pi}{1+k}, \arccos(\kappa)\}$}   & $\mathsf{V}(t_f)\cdot [\mathsf{X}(t_x) \cdot \mathsf{V}(t_v)]^{k} \cdot \mathsf{X}(t_i), \ k \geq 2$  & \multirow{4}{*}{\small{$\frac{\pi}{3} < t_x < \pi$}} \\
\hhline{~~-~}
 & &  $\mathsf{V}(t_f)\cdot [\mathsf{X}(t_x) \cdot \mathsf{V}(t_v)]^{k} \cdot \mathsf{X}(t_x) \cdot \mathsf{V}(t_i), \ k \geq 2$ &  \\
\hhline{~~-~}
 & & $\mathsf{X}(t_f)\cdot [\mathsf{V}(t_v) \cdot \mathsf{X}(t_x)]^{k} \cdot \mathsf{V}(t_i), \ k \geq 2$   & \\
\hhline{~~-~}
 & & $\mathsf{X}(t_f)\cdot [\mathsf{V}(t_v) \cdot \mathsf{X}(t_x)]^{k} \cdot \mathsf{V}(t_v) \cdot \mathsf{X}(t_i), \ k \geq 2$ & \\
\hline
\end{tabular}
}
\caption[Admissible structures of time-optimal sequences, case $t > 0, \ \kappa < \cos(\alpha)$]{\label{T2}Admissible structures of time-optimal sequences, case $t > 0, \ \kappa < \cos(\alpha)$.}
\end{table}

\begin{table}[H]
\centering
\resizebox{\columnwidth}{!}{%
\begin{tabular}{|p{6pt}|c|l|c|c|c|}
\hline
$n$&$\alpha$ range&\hspace{2cm} type & signs & internal angle & outer angles\\
\hline 
\multirow{3}{*}{3} & \multirow{11}{*}{$\alpha < \pi$} & $\mathsf{V}(t_f) \mathsf{X}(t_x)  \mathsf{V}(t_i)$ & $\sgn(t_f) \neq \sgn(t_i)$ & $|t_x| < 2\arccos\left(\frac{|\cos(\alpha)|}{\kappa + |\cos(\alpha)|}\right)$ &  \multirow{3}{*}{}\\
\hhline{~~---~}
 & & $\mathsf{V}(t_f) \mathsf{X}(t_x)  \mathsf{V}(t_i)$ & $\sgn(t_f) = \sgn(t_i)$ & $|t_x| < 2\arccos\left(\frac{\cos(\alpha)-\kappa}{\cos(\alpha)+\kappa}\right)$ &\\
\hhline{~~---~}
 & & $\mathsf{X}(t_f) \mathsf{V}(t_v)  \mathsf{X}(t_i)$ & $\sgn(t_f) \neq \sgn(t_i)$ & $|t_v| < 2\arccos\left(\frac{\kappa |\cos(\alpha)|}{1+\kappa|\cos(\alpha)|}\right)$ &\\
\hhline{-~----}
\multirow{4}{*}{4} &   & $\mathsf{V}(t_f) [\mathsf{X}(t_x)  \mathsf{V}(t_v)]  \mathsf{X}(t_i)$ & $\{+,+,-,-\}$ & \multirow{4}{*}{$|t_x| < 2\arccos\left(\frac{|\cos(\alpha)|}{\kappa + |\cos(\alpha)|} \right)$} &  \multirow{4}{*}{$\begin{cases} |t_{x,\scriptsize{\textrm{out}}}|<\pi-|t_x| \\
|t_{v,\scriptsize{\textrm{out}}}|<\pi-|t_v| \\
|t_{x,\scriptsize{\textrm{out}}}| + |t_{v,\scriptsize{\textrm{out}}}| < \frac{2\pi}{3} \end{cases}$}\\
\hhline{~~--~~}
 & & $\mathsf{V}(t_f) [\mathsf{X}(t_x)  \mathsf{V}(t_v)]  \mathsf{X}(t_i)$ & $\{-,+,+,-\}$ &  &\\
 \hhline{~~--~~}
  & & $\mathsf{X}(t_f) [\mathsf{V}(t_v)  \mathsf{X}(t_x)]  \mathsf{V}(t_i)$ & $\{+,+,-,-\}$ &  &\\
 \hhline{~~--~~}
  & & $\mathsf{X}(t_f) [\mathsf{V}(t_v)  \mathsf{X}(t_x)]  \mathsf{V}(t_i)$ & $\{-,+,+,-\}$ &  &\\
\hhline{-~----}
\multirow{4}{*}{5} &  & $\mathsf{V}(t_f) [\mathsf{X}(t_x)  \mathsf{V}(t_v)]  \mathsf{X}(t_x) \mathsf{V}(t_i)$ & $\{+,+,-,-,+\}$ & \multirow{2}{*}{$|t_x| < \frac{\pi}{2}$} &  \multirow{4}{*}{$\begin{cases} |t_{x,\scriptsize{\textrm{out}}}| < \textrm{min}\{|t_x|, \frac{\pi}{3}\} \\ |t_{v,\scriptsize{\textrm{out}}}| < \textrm{min}\{|t_v|, \frac{\pi}{3}\} \end{cases}$}\\
\hhline{~~--~~}
 & & $\mathsf{V}(t_f) [\mathsf{X}(t_x)  \mathsf{V}(t_v)]  \mathsf{X}(t_x) \mathsf{V}(t_i)$ & $\{-,+,+,-,+\}$ &  &\\
\hhline{~~---~}
 & & $\mathsf{X}(t_f) [\mathsf{V}(t_v)  \mathsf{X}(t_x)]  \mathsf{V}(t_v)   \mathsf{X}(t_i)$ & $\{+,+,-,-,+\}$ & \multirow{2}{*}{$|t_x| < 2\arccot\left(\frac{1}{\kappa}\right)$} &\\
\hhline{~~--~~}
 & & $\mathsf{X}(t_f) [\mathsf{V}(t_v)  \mathsf{X}(t_x)]  \mathsf{V}(t_v)   \mathsf{X}(t_i)$ & $ \{+,-,-,+,+\}$ &  &\\
\hhline{------}
\multirow{6}{*}{\hspace{-2pt}$\infty$} &
\multirow{2}{*}{$\arccos(\kappa) < \alpha < \pi$}   & $\mathsf{V}(t_f) \mathsf{Q}(t_Q)  \mathsf{X}(t_i)$ & $\{+,+,+\}$ & \multirow{2}{*}{} & \multirow{6}{*}{} \\
\hhline{~~--~~}
 & &  $\mathsf{X}(t_f) \mathsf{Q}(t_Q)  \mathsf{V}(t_i)$ & $ \{+,+,+\}$ & & \\
\hhline{~----~}
 & \multirow{4}{*}{$\alpha < \pi - \arccos(\kappa)$}  & $\mathsf{V}(t_f) \mathsf{P}(t_P)  \mathsf{V}(t_i)$ & $\sgn(t_f) \neq \sgn(t_i)$  & $|t_p| < 2\arccos\left(   \frac{\kappa}{1+\kappa}\right)$ &  \\
\hhline{~~---~}
 & & $\mathsf{X}(t_f) \mathsf{P}(t_P)  \mathsf{X}(t_i)$ & $ \sgn(t_f) \neq \sgn(t_i)$ & $|t_p| < \frac{2\pi}{3}$ &\\
\hhline{~~---~}
 & & $\mathsf{V}(t_f) \mathsf{P}(t_P)  \mathsf{X}(t_i)$ & $ \sgn(t_f) \neq \sgn(t_i)$ & \multirow{2}{*}{} &\\
\hhline{~~--~~}
 & & $\mathsf{X}(t_f) \mathsf{P}(t_P)  \mathsf{V}(t_i)$ & $ \sgn(t_f) \neq \sgn(t_i)$ & &\\
\hline
\end{tabular}
}
\caption[Admissible structures of time-optimal sequences, case $t \gtrless 0$]{\label{T3}Admissible structures of time-optimal sequences, case $t \gtrless 0$; the shown sign combinations are relative, that is, the sequences remain admissible under a global sign change.}
\end{table}

\section{Applications}
\label{sec:apps}
The restricted control set that was studied in the preceding Sections is of relevance in many electron-nuclear spin systems exhibiting anisotropic hyperfine couplings, for example: a $^{13}$C proximal to an NV center in diamond~\cite{Jelezko05, Childress06}; a proton coupled to a free electron in malonic acid~\cite{Hodges08,Khaneja07,Mitrikas10}; $^{31}$P in P donors in Si~\cite{Morton08}; N in buckyballs~\cite{Morton06}; and other quantum compounds studied in Nuclear Magnetic Resonance~\cite{Assemat10}. For such systems, the nuclear evolution can be steered via the switching of the actuator-electronic spin, in a generally faster and noise-free way, as compared to the direct addressing of the nuclear spin. 

Specifically for the coupled qubits in diamond, we have recently shown~\cite{Aiellounpublished} that this actuator protocol for driving the $^{13}$C nuclear spin is in general advantageous over radio-frequency direct driving, especially for external magnetic fields in the range $B_0 \sim 250-500$G, and bare nuclear Rabi frequencies $\Omega \lesssim 2\pi \cdot 20$kHz such as those which are usually obtained with modest amplifiers.  

Additionally, the same control set is used to model machine motion such as satellite reorientation~\cite{Boscain05,Trelat12}, so that we believe our results will be of interest to the robotics community as well.

\begin{figure}[t]
\centering
\includegraphics[width=.27\columnwidth]{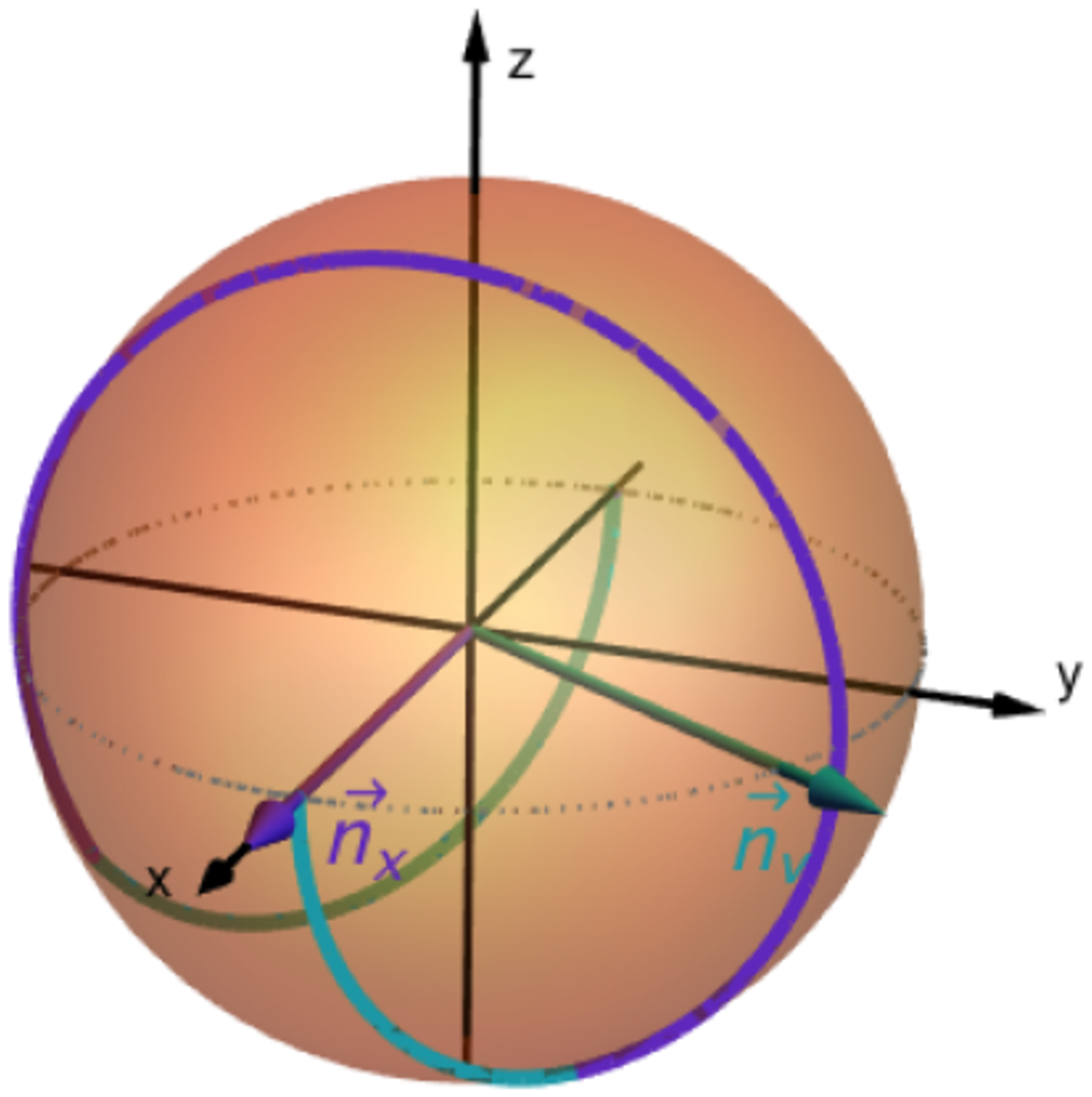}\qquad
\includegraphics[width=.27\columnwidth]{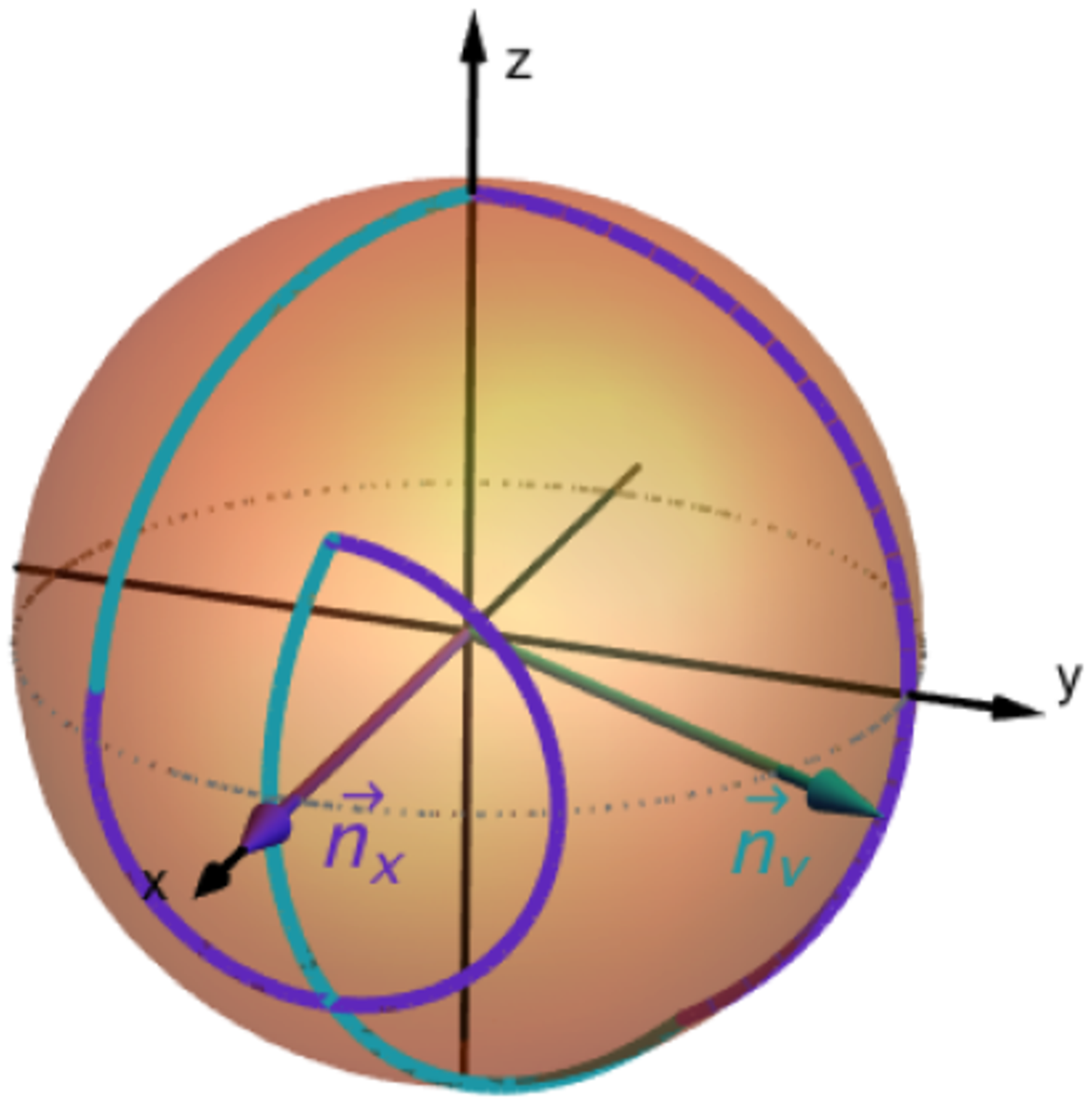}\qquad
\includegraphics[width=.27\columnwidth]{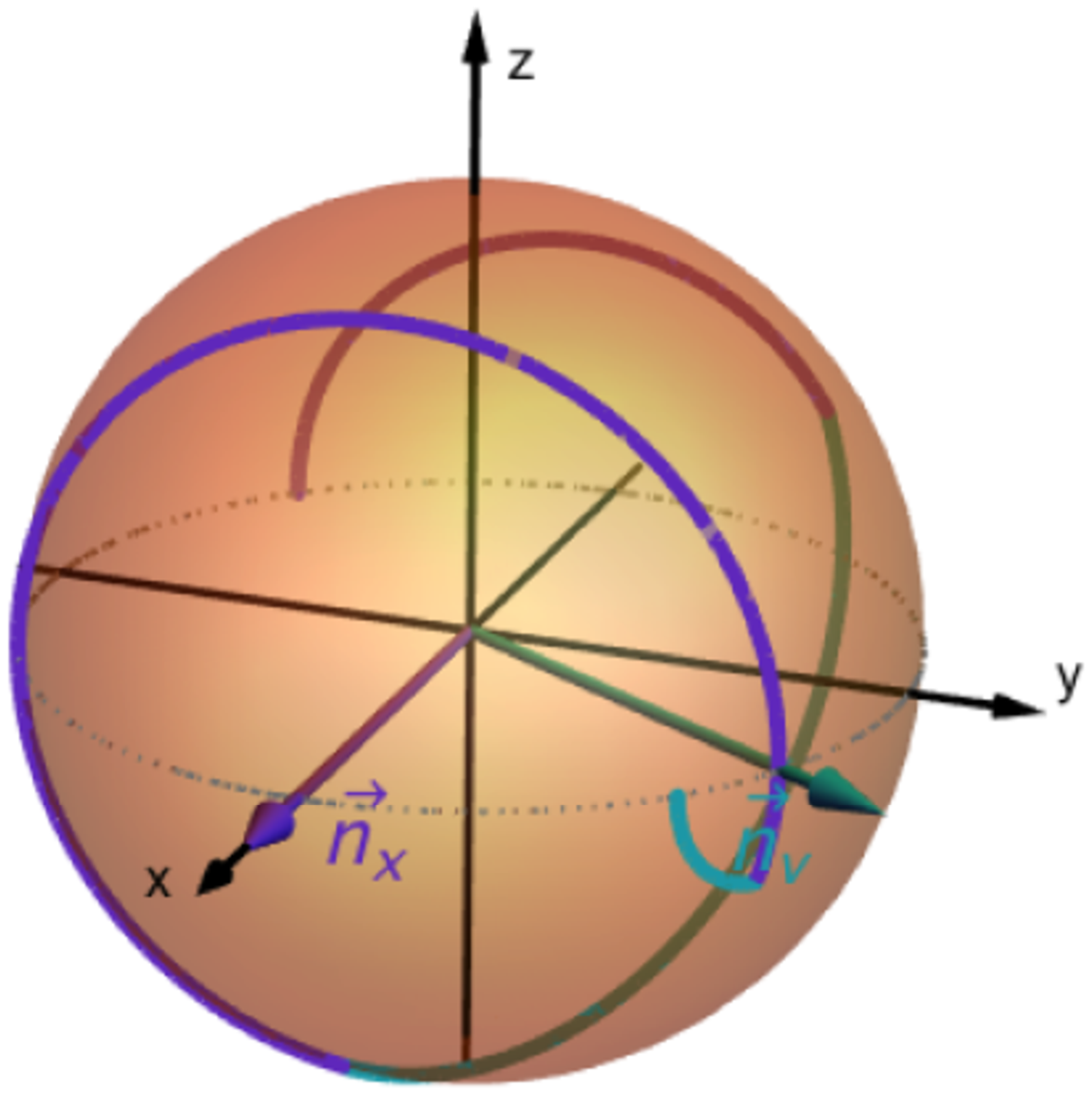}\vspace{12pt}
\includegraphics[width=.9\columnwidth]{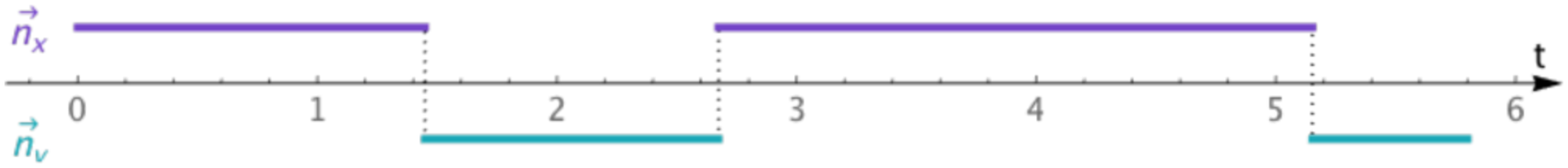}
\caption{Time-optimal control solution to obtain a $\pi$ rotation about the $\vec{n}_z$ axis for $\alpha=\pi/3$ and $\kappa=1/4$. The time-optimal solution has four control concatenations, as represented on the sphere in terms of the rotation angles, and on the bottom plot in terms of normalized times. It can be applied to any initial state $\vec r(0)$: left, $\vec r(0)=\vec{n}_x$; center, $\vec r(0)=\vec{n}_z$ (so no net rotation is obtained); right, $\vec r(0)=(\vec{n}_x+\vec{n}_y)/\sqrt2$.}
\label{fig:Example14}
\end{figure}

\begin{figure}
\centering
\includegraphics[width=.27\columnwidth]{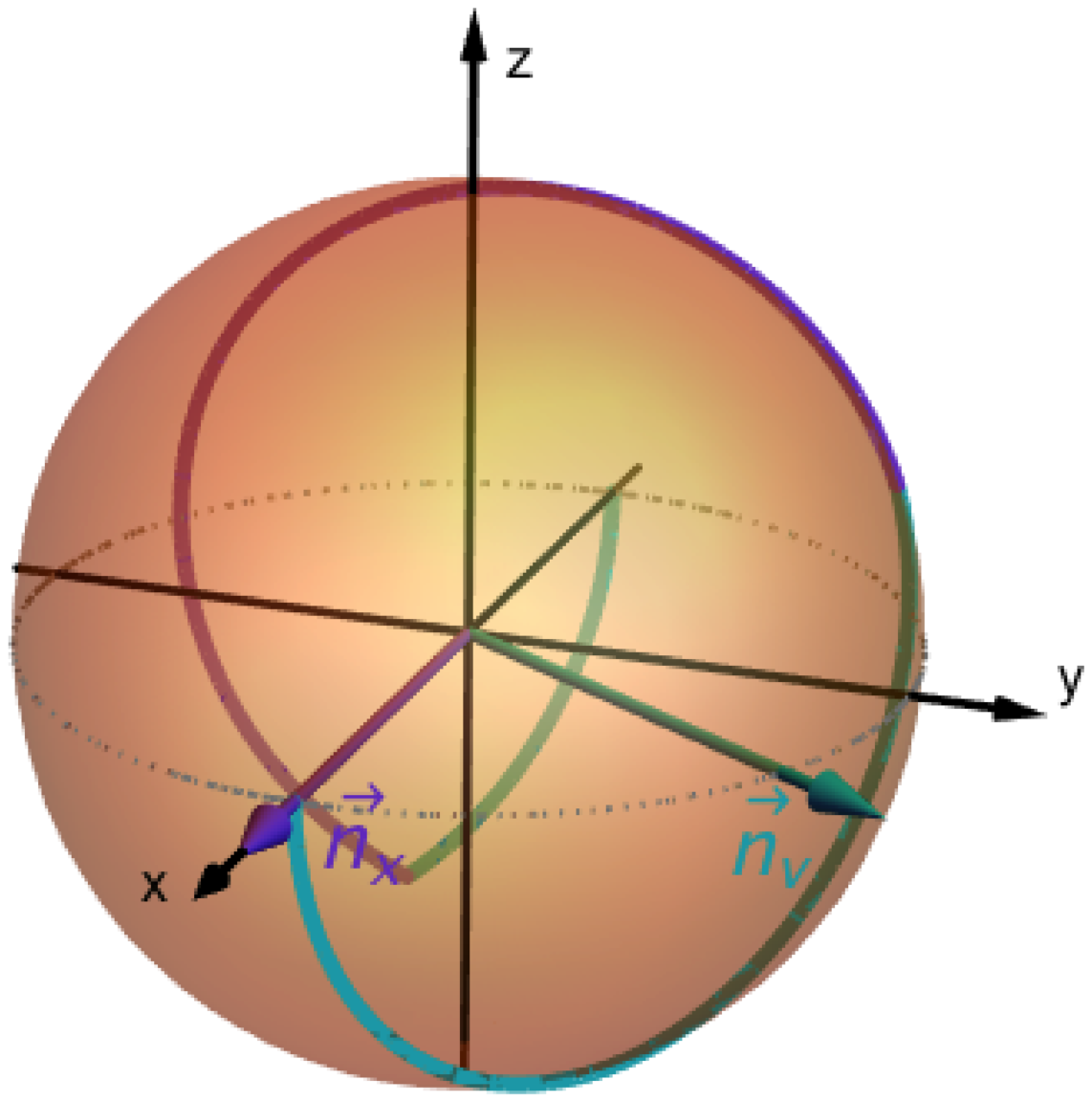}\qquad
\includegraphics[width=.27\columnwidth]{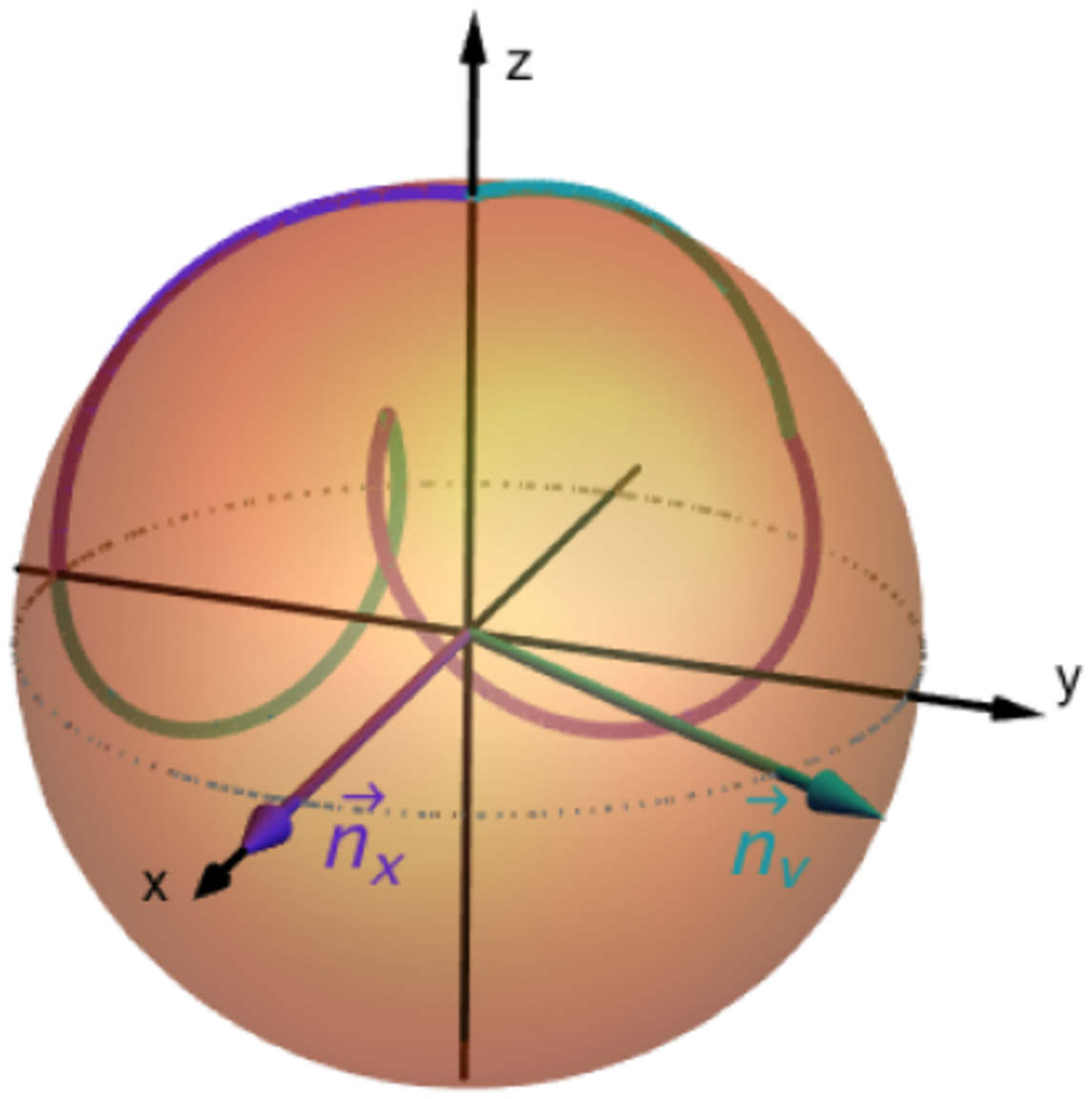}\qquad
\includegraphics[width=.27\columnwidth]{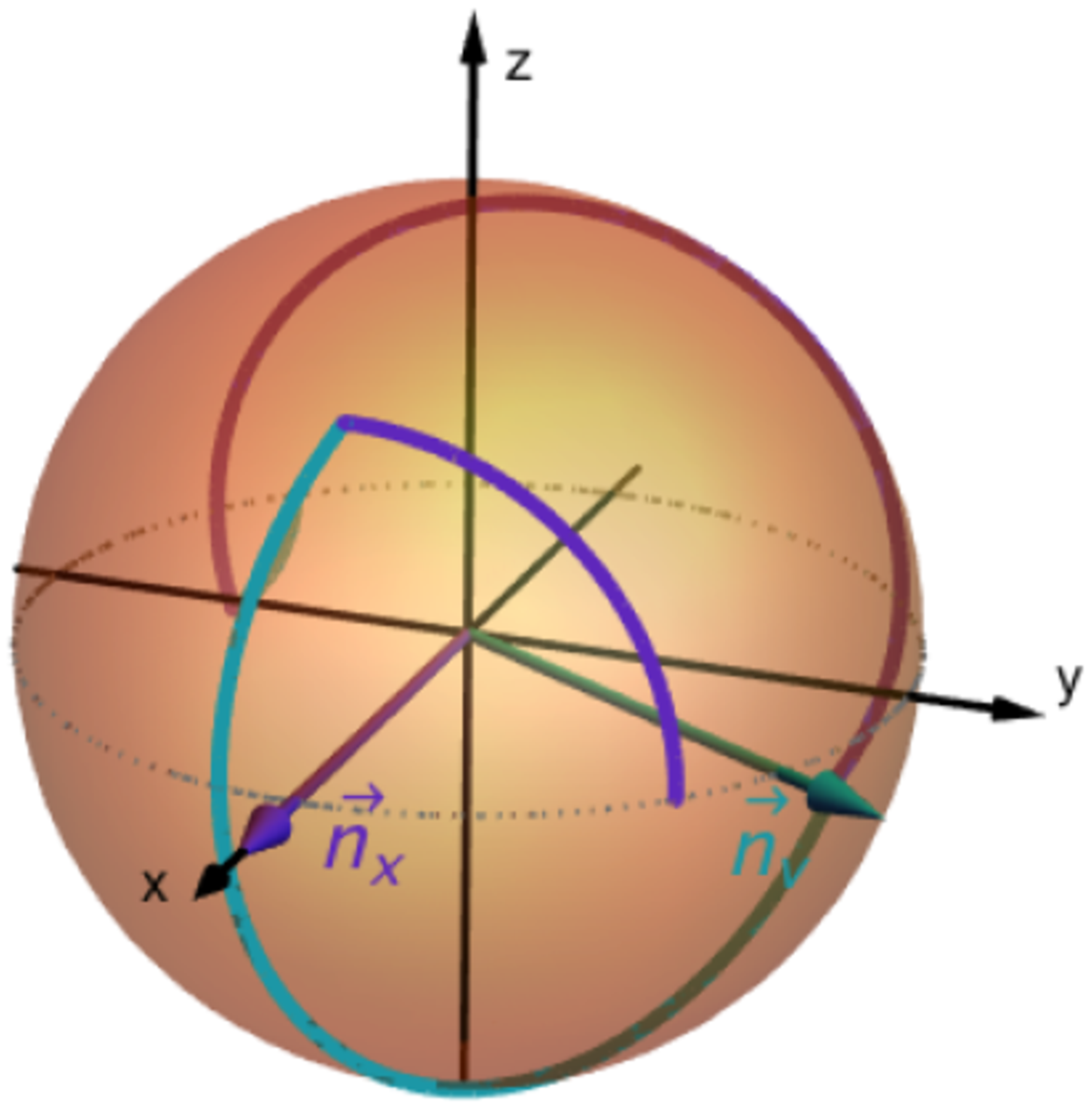}\vspace{12pt}
\includegraphics[width=.8\columnwidth]{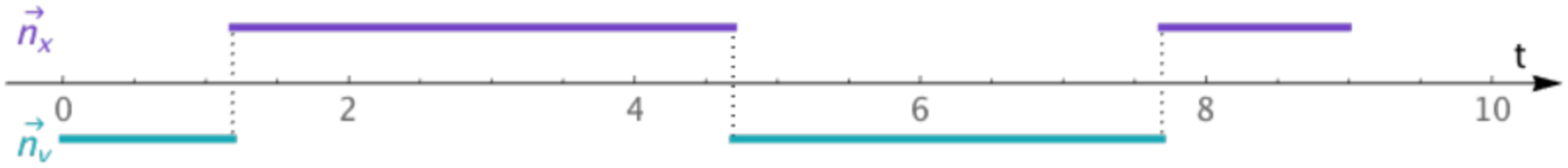}
\caption{Time-optimal control solution to obtain a $\pi$ rotation about the $\vec{n}_z$ axis for $\alpha=\pi/3$ and $\kappa=3/4$. The time-optimal solution is represented in a similar way to Figure~\ref{fig:Example14}.  It can be applied to any initial state $\vec r(0)$: left, $\vec r(0)=\vec{n}_x$; center, $\vec r(0)=\vec{n}_z$ (so no net rotation is obtained); right, $\vec r(0)=(\vec{n}_x+\vec{n}_y)/\sqrt2$.}
\label{fig:Example34}
\end{figure}

Two examples of time-optimal solutions, as found by a numerical search constrained by the derived necessary conditions, are depicted in Figures~\ref{fig:Example14},~\ref{fig:Example34}.

\section{Conclusion}
\label{sec:conc}
In conclusion, we have addressed the problem of time-optimal generation of $SU(2)$ unitaries through concatenations of elementary rotations
about two non-parallel, and generally non-orthogonal, axes. We have algebraically derived the necessary general structure of time-optimal sequences, and shown that such sequences are described solely in terms of three independent parameters representing rotation angles, and the total number of rotations $n$. Bounds for such parameters were found, as a function of the angle between the rotation axes, $\alpha$, and a parameter describing the difference in effective implementation time, or experimental cost, of the rotations, $\kappa$. Given the experimentally relevant $\alpha, \kappa$, in certain cases we can also predict the maximal number of rotations $n$ in a time-optimal sequence. 

Our method maps an optimization problem involving differential equations into a much simpler, algebraic linear problem. While our analysis starts from abstract mathematical results in optimal control theory, we go beyond previous literature in providing the
experimental physicist with a general set of instructions to find the time-optimal operations in a large
set of realistic experimental conditions. While these instructions are in general not sufficient to single out the
time-optimal sequence for the desired unitary, they provide a very
powerful set of rules that constrains the structure of time-optimal
solutions so strongly, that the solution can be found
through a simple numerical search.

The key interest of our results stems from their wide applicability to
quantum systems with a restricted control set. In particular, we envision fast unitary control of a nuclear spin by switching the spin states of an electronic spin, in the case of anisotropic hyperfine interaction. This setting occurs, for example, in a proximal $^{13}$C coupled to a NV center in diamond. 

Furthermore, outside quantum science, the very general control problem we address will be of interest in diverse fields of physics and engineering, for instance robotics; the accessible approach we employ, and the power of the general results and insights into the structure of time-optimal sequences it provides, are bound to become an invitation to the physicist un-initiated in theoretical control methods.

\section{Acknowledgments}
This work was supported in part by the U.S. Air Force Office of Scientific Research through the Young Investigator Program. C.D.A acknowledges support from Schlumberger. The authors would like to thank Seth Lloyd for discussions, and for pointing out reference~\cite{Billig13}; exchanges with Ugo Boscain and Domenico D'Alessandro are also gratefully acknowledged.
\section*{References}

\bibliography{main}

\end{document}